\documentclass[prd,aps,twocolumn,preprintnumbers, showpacs, nofootinbib,superscriptaddress,notitlepage]{revtex4-1}
\usepackage{amsmath,graphicx,epsfig,amssymb}
\usepackage{inputenc}
\usepackage[usenames]{color}
\usepackage{ulem} 
\usepackage{bigstrut}
\usepackage{slashed}
\usepackage{multirow}
\usepackage{subfigure}
\usepackage{dsfont}
\usepackage{xcolor}
\usepackage{hyperref}
\usepackage{comment}
\usepackage{amsthm}
\usepackage[capitalise,nameinlink]{cleveref}
\crefname{equation}{Eq.}{Eqs.}
\usepackage{hyperref} 
\hypersetup{
    colorlinks = true, 
    linkcolor = blue, 
    citecolor = blue, 
    urlcolor = black 
}
\makeatletter
\let\oldtagform@\tagform@
\renewcommand\tagform@[1]{%
  \hypertarget{eq:#1}{\oldtagform@{#1}}%
}
\renewcommand{\eqref}[1]{%
  \textup{\hyperref[eq:#1]{\oldtagform@{\ref*{eq:#1}}}}%
}
\makeatother

\allowdisplaybreaks

\begin{document}

\title{Approaches to the Inverse Fourier Transformation with Limited and Discrete Data}

\author{Yu-Fei Ling}
\affiliation{Key Laboratory of Atomic and Subatomic Structure and Quantum Control (MOE), 
Guangdong Basic Research Center of Excellence for Structure and Fundamental Interactions of Matter, 
Institute of Quantum Matter, South China Normal University, Guangzhou 510006, China}
\affiliation{Guangdong-Hong Kong Joint Laboratory of Quantum Matter, 
Guangdong Provincial Key Laboratory of Nuclear Science, Southern Nuclear Science Computing Center, 
South China Normal University, Guangzhou 510006, China}

\author{Min-Huan Chu}
\affiliation{Faculty of Physics and Astronomy, Adam Mickiewicz University, ul.\ Uniwersytetu Pozna\'nskiego 2, 61-614 Pozna\'{n}, Poland}

\author{Jian Liang}
\affiliation{Key Laboratory of Atomic and Subatomic Structure and Quantum Control (MOE), 
Guangdong Basic Research Center of Excellence for Structure and Fundamental Interactions of Matter, 
Institute of Quantum Matter, South China Normal University, Guangzhou 510006, China}
\affiliation{Guangdong-Hong Kong Joint Laboratory of Quantum Matter, 
Guangdong Provincial Key Laboratory of Nuclear Science, Southern Nuclear Science Computing Center, 
South China Normal University, Guangzhou 510006, China}

\author{Jun Hua}
\email{Corresponding author: junhua@scnu.edu.cn}
\affiliation{Key Laboratory of Atomic and Subatomic Structure and Quantum Control (MOE), 
Guangdong Basic Research Center of Excellence for Structure and Fundamental Interactions of Matter, 
Institute of Quantum Matter, South China Normal University, Guangzhou 510006, China}
\affiliation{Guangdong-Hong Kong Joint Laboratory of Quantum Matter, 
Guangdong Provincial Key Laboratory of Nuclear Science, Southern Nuclear Science Computing Center, 
South China Normal University, Guangzhou 510006, China}

\author{Ao-Sheng Xiong}
\affiliation{Frontiers Science Center for Rare Isotopes, and School of Nuclear Science and Technology, Lanzhou University, Lanzhou 730000, China}

\author{Qi-An Zhang}
\email{Corresponding author: zhangqa@buaa.edu.cn}
\affiliation{School of Physics, Beihang University, Beijing 102206, China}

\begin{abstract}
We investigate several approaches to address the inverse problem that arises in the limited inverse Fourier transform (L-IDFT) of quasi-distributions. The methods explored include Tikhonov regularization, the Backus–Gilbert method, the Bayesian approach with Gaussian Random Walk (GRW) prior, and the feedforward artificial neural networks (ANNs). We evaluate the performance of these methods using both mock data generated from toy models and real lattice data from quasi distribution, and further compare them with the physics-driven $\lambda$-extrapolation approach. Our results demonstrate that the L-IDFT constitutes a moderately tractable inverse problem Except for the Backus–Gilbert method, all the other approaches are capable of correctly reconstructing the quasi-distributions in momentum space. In particular, the Bayesian approach with GRW and the feedforward ANNs yield more stable and accurate reconstructions. Based on these investigations, we conclude that, for a given L-IDFT problem, selecting an appropriate reconstruction method according to the input data and carefully assessing the potential systematic uncertainties are essential for obtaining reliable results.
\end{abstract}

\maketitle

\section{Introduction}\label{sec:introduction}
The determination of partonic structure from first principles remains one of the central challenges in quantum chromodynamics (QCD). Lattice QCD provides a systematically improvable approach to nonperturbative structures. However, the Euclidean formulation of lattice prevents a direct computation of correlation functions defined on the light cone, such as parton distribution functions (PDFs), light-cone distribution amplitudes (LCDAs), generalized parton distributions (GPDs), and transverse-momentum dependent distributions (TMDs). The traditional operator product expansion (OPE) techniques can only access a limited number of Mellin moments of structure functions. 
Recent years, the large-momentum effective theory (LaMET)~\cite{Ji:2013dva,Ji:2014gla} has opened a promising avenue by relating Euclidean “quasi-distributions” computed in large momenta boosted hadrons to their corresponding light-cone counterparts through factorization and perturbative matching~\cite{Zhang:2017bzy,Zhang:2020gaj,Holligan:2023rex,Hua:2020gnw,LatticeParton:2022zqc,Baker:2024zcd,Cloet:2024vbv,Xiong:2013bka,Alexandrou:2016eyt,Chen:2017mie,Zhang:2017zfe,Xu:2018mpf,Liu:2018hxv,Wang:2019msf,Zhang:2019qiq,Chen:2020ody,Ji:2020brr,Ji:2020ect,LatticeParton:2020uhz,Lin:2020rxa,Bhattacharya:2021moj,Gao:2021hxl,LatticePartonLPC:2021gpi,Li:2021wvl,Deng:2022gzi,Gao:2022iex,Gao:2022uhg,LatticeParton:2022xsd,LatticePartonCollaborationLPC:2022myp,LatticePartonLPC:2022eev,Zhang:2022xuw,Zhu:2022bja,Deng:2023csv,Ji:2023pba,LatticeParton:2023xdl,Zhao:2023ptv,Liu:2023onm,Avkhadiev:2024mgd,Good:2024iur,Han:2024cht,Han:2024min,Holligan:2024umc,Holligan:2024wpv,Ji:2024hit,Wang:2024wwa,LatticeParton:2024zko,Zhang:2024omt,Bollweg:2025iol,Wang:2025uap,Ji:2025mvk}. A closely related framework, the pseudo-distribution method~\cite{Radyushkin:2017cyf,Orginos:2017kos}, instead makes use of coordinate-space correlation functions evaluated at multiple hadron momenta, and reconstructs the light-cone distribution by matching and fitting the model functions in momentum space\cite{Radyushkin:2017lvu,Zhang:2018ggy,Karpie:2018zaz,Joo:2019jct,Joo:2019bzr,HadStruc:2021qdf,Karpie:2021pap,Bhat:2022zrw,Kovner:2024pwl,Bhattacharya:2024qpp,HadStruc:2024rix,NieMiera:2025inn}.

An essential numerical step in both the LaMET and pseudo-distribution approaches is the reconstruction of momentum-fraction distributions from nonlocal matrix elements computed in Euclidean lattice QCD. This problem is mathematically equivalent to performing a limited inverse discrete Fourier transform (L-IDFT), since the data are available only at finite and discrete non-local separations, and are inevitably contaminated by statistical noise. Recently, several works have addressed the problem of L-IDFT from different perspectives~\cite{Karpie:2019eiq,Dutrieux:2024rem,Chen:2025cxr,Xiong:2025obq,Dutrieux:2025jed,Medrano:2025cmg}. In the hybrid scheme~\cite{Ji:2020brr,Chen:2025cxr}, it was shown that if lattice data can be obtained at sufficiently large nonlocal separations, one may employ physically motivated extrapolations to effectively avoid the inverse problem. In the recent study \cite{Xiong:2025obq} for ill-posedness in limited discrete fourier inversion and its regularization, we provided a rigorous mathematical proof that the L-IDFT satisfies the Hadamard criteria of existence and uniqueness, but fails the stability condition: small perturbations in the input can lead to exponentially amplified deviations in the reconstructed distribution. This formulates the L-IDFT as a genuine inverse problem.  We further demonstrated that it falls into the class of moderately tractable ill-posed problems, implying that established methods from the theory of inverse problems can be successfully applied to stabilize the reconstruction.

A natural question then arises: for the L-IDFT inverse problem, which methods are most suitable for achieving a stable reconstruction? From the perspective of inverse problems, a variety of complementary strategies have been proposed and successfully applied in related contexts~\cite{Li:2020ejs,Li:2020xrz,Li:2021gsx,Li:2022qul,hansen1994regularization,mueller2012linear,kabanikhin2011inverse,lesnic2021inverse}. In this work, we explore four representative approaches. The Tikhonov regularization method ~\cite{Xiong:2022uwj,Kirsch2021,vogel2002computational,calvetti2000tikhonov,benning2018modern} stabilizes the inversion by penalizing large fluctuations, with well-established theoretical guarantees. The Backus–Gilbert method~\cite{backus-gilbert} constructs optimized linear estimators by minimizing the variance of a resolution kernel, thereby achieving a transparent balance between bias and variance. The Bayesian approach~\cite{PhysRevLett.111.182003,Bayesian_Inverse_problem_application,stuart2010inverse,knapik2011bayesian} (implemented with a Gaussian Random Walk (GRW) prior and Markov Chain Monte Carlo (MCMC) sampling in this work), provides a probabilistic framework for incorporating prior information and propagating uncertainties, and has been widely applied to spectral-function reconstruction. Finally, the feedforward artificial neural networks (ANNs)~\cite{2023arXiv231014459G, ZHANG2025108544, Chouzenoux_2025,10.7551/mitpress/7132.001.0001} offer a nonparametric and flexible approximation scheme, which can implicitly regularize ill-posed problems and have demonstrated strong performance in global parton distribution analyses.

This work presents a systematic comparison of several representative approaches for solving the limited inverse discrete Fourier transform problem. We develop a unified testing strategy to evaluate their performance, employing both controlled toy models and real lattice data relevant to LaMET studies of light cone distribution amplitudes (LCDAs). Our results show that, in most cases, the Tikhonov regularization, the Bayesian approach with GRW, and the feedforward artificial neural network can successfully reconstruct the quasi-distributions in momentum space. Among them, the Bayesian and ANN methods exhibit superior stability and accuracy. Our study and comparisons further demonstrate that the inverse Fourier transform problem, unlike those encountered in spectral analyses or in reconstructing distributions from OPE moments, is a moderately tractable ill-posed problem. When lattice data of sufficient quality are available at large nonlocal separations, this problem can indeed be effectively addressed. These investigations thus provide valuable methodological insights for future lattice QCD extractions of parton physics.
The paper is organized as follows: In Sec.~\ref{sec_framework}, we provide a brief overview of the formulation of ill-posed inverse problems and introduce four approaches for addressing the limited inverse discrete Fourier transform: Tikhonov regularization, the Backus–Gilbert method, the Bayesian approach with GRW, and the feedforward artificial neural networks. In Sec.~\ref{sec_numresultes}, we present numerical tests of these four methods using two distinct toy models and one set of real lattice data. Finally, Sec.~\ref{sec_summary} provides a summary and concluding remarks of the present study.

\section{Framework}\label{sec_framework}
\subsection{Definition of Ill-posed Inverse Problems}

To establish a precise definition of inverse problems, it is instructive to 
first clarify the notion of a forward problem. In scientific and engineering 
research, a fundamental objective is to predict a set of observable outputs 
$g \in G$, from a given collection of input $f \in F$ by a prescribed mathematical relation or operator mapping $K: F \to G$. Therefore, a forward relation satisfies:
\begin{align}
    K(f) = g. \label{eq:forward}
\end{align}
Here, $F$ and $G$ are typically taken to be normed spaces or Hilbert spaces, depending on the functional structure of the problem. 

Based on the above definition of forward problems, the notion of inverse problems can be naturally introduced. In situations where the observables $g \in G$ are known, either fully or partially, one attempts to recover the inaccessible parameters $f \in F$ by means of an inverse mapping $K^{-1}$. Such problems are common in contexts where physical quantities are difficult to measure directly in a given representation or space. 
In lattice QCD, a well known example arises in reconstructing the energy spectrum in momentum space from correlation functions computed in discrete coordinate space. Similarly, within the frameworks of LaMET and the pseudo-distribution approach, one can only compute quasi-distributions or pseudo-distributions in coordinate space with finite extent. Recovering the corresponding distributions in momentum space requires applying a limited inverse discrete Fourier transform, which poses analogous challenges.

For most inverse problems, obtaining a stable and reliable solution is highly nontrivial, and in some cases even impossible. A crucial first step in assessing the nature of an inverse problem is to examine Hadamard’s well-posedness criteria: \textbf{existence}, \textbf{uniqueness}, and \textbf{stability}~\cite{Kirsch2021}. Violation of any of these conditions renders the problem ill-posed; moreover, the degree of ill-posedness generally increases as more conditions are violated or as the violation of a single condition becomes more severe, making the reconstruction progressively more difficult or even unsolvable.
In our previous work~\cite{Xiong:2025obq}, we demonstrated that the L-IDFT satisfies the criteria of existence and uniqueness but fails to meet stability, identifying it as a moderately ill-posed yet tractable problem. Consequently, in this work, we focus on strategies to effectively mitigate the inherent instability.

\subsection{Limited Inverse Discrete Fourier Transform}
Within the frameworks of LaMET or the pseudo-distribution approach, the equal-time correlators in coordinate space and  momentum space are related through a Fourier transform. We take the quasi-distribution as example,
\begin{align}
    g(\lambda)=\int dxe^{-i\lambda x}f(x), \label{eq_intFT}
\end{align}
where $\lambda=z\cdot p_z$ represents the dimensionless variable corresponding to the Euclidean non-local separation. $g(\lambda)$ denotes the renormalized matrix element of a nonlocal equal-time correlator carrying a large momentum, while $f(x)$ stands for the corresponding quasi-distribution in momentum space. 

In lattice numerical simulations, the integral in Eq.~(\ref{eq_intFT}) is discretized into a summation. Furthermore, both $g(\lambda)$ and $f(x)$ can be discretized and represented as vectors, such that Eq.~(\ref{eq_intFT}) can be reformulated as a mapping, in which $K$ denotes the corresponding transformation matrix:
\begin{align}
    &\boldsymbol{g}_k=\sum_{j}\boldsymbol{K}_{kj}\boldsymbol{f}_j,  \notag \\ 
   &\boldsymbol{K}_{kj}= e^{-i \lambda_k x_j}\Delta x, \notag \\ 
   &\boldsymbol{f}=[f(x_0),f(x_1),\cdots,f(x_j)\cdots,f(x_m)]^T, \notag \\
   &\boldsymbol{g}=[g(\lambda_0),g(\lambda_1),\cdots,g(\lambda_k)\cdots,g(\lambda_n)]^T. \label{eq_mapFT}
\end{align}
It should be noticed that, the quasi distribution $f(x)$ in momentum space is pure real according to physical requirement, while the $K_{kj}$ and $g(\lambda)$ are complex. 
In practical lattice simulations, the observable $g(\lambda)$ inevitably contains statistical uncertainties. For convenience of subsequent discussions, we introduce $\boldsymbol{\delta}$ to denote a Gaussian-distributed error, and use $\boldsymbol{g}_{\text{true}}$ and $\boldsymbol{f}_{\text{true}}$ to represent the exact results in coordinate space and momentum space, respectively:
\begin{align}
   \boldsymbol{g}^{\boldsymbol{\delta}}&=\boldsymbol{K}\boldsymbol{f}_{\text{true}}+\boldsymbol{\delta}
   =\boldsymbol{g}_{\text{{true}}}+\boldsymbol{\delta}. \label{eq:datamodeling}
\end{align} 

In general, one can obtain the momentum-space solution from coordinate-space results through a well-defined inverse Fourier transformation. However, in the case of quasi-distributions calculated in lattice, only a finite and discrete set of $\lambda$ values is available, which turns the problem into a limited inverse discrete Fourier transform(L-IDFT):
\begin{align}
   f(x)&=\int_{-\infty}^{\infty}\frac{d\lambda}{2\pi} e^{i\lambda x}g(\lambda) \to \sum^{\lambda_{\text{cut}}}_{-\lambda_{\text{cut}}}\frac{\Delta\lambda}{2\pi}e^{i\lambda x}g(\lambda). \label{eq:inverseFT}
\end{align}
The truncation of $\lambda$ inevitably leads to the loss of tail information in coordinate space, which corresponds to the loss of high-frequency components in momentum space. Such information loss constitutes a typical feature of inverse problems and, as we will see in the subsequent analysis, gives rise to instabilities. From this perspective, if one could effectively reconstruct the missing information in coordinate space, the ill-posed inverse problem could be circumvented. This consideration also explains why physics-driven $\lambda$-extrapolation has proven effective in most LaMET studies~\cite{LatticeParton:2022zqc,Chen:2025cxr}.

Another alternative approach to reconstructing $f(x)$ is to start from Eq.~(\ref{eq_mapFT}) and directly invert the transformation matrix $K$:
\begin{align}
   \boldsymbol{f}=(\boldsymbol{K}^{\dagger}\boldsymbol{K})^{-1}\boldsymbol{K}^{\dagger}\boldsymbol{g}. \label{eq_gen_inverse}
\end{align}
In cases where $\boldsymbol{K}$ is not a square matrix, we employ the pseudo-inverse to represent the inversion. The L-IDFT problem in this work satisfies the conditions of existence and uniqueness, but lacks stability. Within this formulation, the instability can be clearly revealed through singular value decomposition (SVD). Since both $\boldsymbol{K}$ and $\boldsymbol{g}$ in Eq.~(\ref{eq_gen_inverse}) are complex , we will, for simplicity, restrict the entire subsequent discussion to their real parts:
\begin{align}
   \boldsymbol{K}_{\text{re}}&=\boldsymbol{U}\boldsymbol{\Sigma}\boldsymbol{V}^T \notag \\
   &=\sum_{i=1}^{n}\boldsymbol{u}_i\sigma_i\boldsymbol{v}_i^T, \quad n=\text{min}(n_\lambda,n_x),
\end{align}
where 
\begin{align}
   \boldsymbol{U}&=[\boldsymbol{u}_1,\cdots,\boldsymbol{u}_n]\in \mathbb{R}^{n_{\lambda}\times n},\quad \boldsymbol{U}^T\boldsymbol{U}=\boldsymbol{I}_{n\times n}, \notag \\
   \boldsymbol{V}&=[\boldsymbol{v}_1,\cdots,\boldsymbol{v}_n]\in \mathbb{R}^{n_x\times n}, \quad \boldsymbol{V}^T\boldsymbol{V}=\boldsymbol{I}_{n\times n}, \notag \\
   \boldsymbol{\Sigma}&=\text{diag}[\sigma_1,\cdots,\sigma_n],\quad \sigma_1\ge\cdots\ge\sigma_n\ge0,
\end{align}
here $\boldsymbol{U}$ and $\boldsymbol{V}$ denote singular vectors with orthonormal columns, while $\boldsymbol{\Sigma}$ is the singular value matrix of $\boldsymbol{K}_{\text{re}}$~\cite{allen1985singular}. Therefore, the formal solution of Eq.~(\ref{eq_mapFT}) takes the form:
\begin{align}
   \boldsymbol{f}=\sum_{i=1}^{n}\frac{\boldsymbol{u}_i^T\boldsymbol{g}}{{\sigma}_i}\boldsymbol{v}_i.
\end{align}
When the data $g$ are contaminated by noise, the reconstructed result becomes:
   \begin{align}
       ||\boldsymbol{f}^{\boldsymbol{\delta}}-\boldsymbol{f}_{\text{true}}||_2^2&=\sum_{i=1}^{n}(\frac{\boldsymbol{u}_i^T(\boldsymbol{g}^{\boldsymbol{\delta}}-\boldsymbol{g}_{\text{true}})}{{\sigma}_i}\boldsymbol{v}_i)^2, \notag \\
       &=\sum_{i=1}^{n}(\frac{\boldsymbol{u}_i^T\boldsymbol{\delta}}{{\sigma}_i}\boldsymbol{v}_i)^2, \label{eq_solution_variation}
   \end{align}
in which $||\cdot||_2$ denotes the 2-norm. Eq.~(\ref{eq_solution_variation}) reveals the direct dependence of error amplification on the singular values, motivating the definition of the condition number $\text{cond}(\boldsymbol{K}_{\text{re}})_2=\sigma_{\text{max}}/\sigma_{\text{min}}=\sigma_1/\sigma_n$. 
As documented in Ref.~\cite{Xiong:2025obq}, the singular value spectrum of $\boldsymbol{K}_{\text{re}}$ exhibits a smallest singular value that approaches zero. This leads to extreme amplification of errors in the solution and a correspondingly large condition number. 
Consequently, for the inverse problem violating stability, even a slight perturbation of $g(\lambda)$ results in large variations in the inverse solution, providing no reliable or useful information. 
A stable solution to such an ill-posed problem can be obtained by several types of mathematical techniques, as will be detailed in following subsections.

\subsection{Tikhonov regularization method} \label{sec:Tik_reg}
Tikhonov regularization is a standard technique for addressing inverse problems. 
It proceeds by augmenting the objective functional with an additional penalty term 
$\alpha \lVert \boldsymbol{\Gamma}\boldsymbol{f} \rVert_2^2$, and subsequently determining the regularized solution $\boldsymbol{f}_{\text{reg}}^{\alpha}$ that minimizes this modified functional~\cite{Xiong:2022uwj}:
\begin{align}
    \boldsymbol{f}_{\text{reg}}^{\alpha}&=\text{arg}\min_{\boldsymbol{f}}L(\boldsymbol{f}), \notag\\
    &=\text{arg}\min_{\boldsymbol{f}} [(\boldsymbol{K}\boldsymbol{f} - \boldsymbol{g})^{T} (\boldsymbol{K}\boldsymbol{f} - \boldsymbol{g}) + \alpha \| \boldsymbol{\Gamma}\boldsymbol{f} \|^2_2], \label{eq:reg}
\end{align}
where $\boldsymbol{\Gamma}$ approximates the derivation operator and $\alpha$ is the regularization parameter. By differentiating the objective functional and setting it to zero $\partial L/{\partial \boldsymbol{f}} =0$, the regularized solution is given as:
\begin{align}
    \boldsymbol{f}_{\text{reg}}^{\alpha}&=(\boldsymbol{K}^{T} \boldsymbol{K}+\alpha \boldsymbol
    {\Gamma}^{T}\boldsymbol{\Gamma})^{-1}\boldsymbol{K}^{T}\boldsymbol{g}.
    \label{show_the_way_to_overcome_the_instability}
\end{align} 
Here $\boldsymbol{\Gamma}$ is typically chosen as identity matrix $\boldsymbol{I}$, which means that the true solution is continuous \cite{hansen1994regularization}. From Eq.~(\ref{show_the_way_to_overcome_the_instability}), it is obvious that the Tikhonov regularization can decrease the condition number of the ill-posed inverse problem, especially for these instability problems.
\begin{itemize}
    \item 
    Original inverse problem:
   \begin{align}
      \boldsymbol{K}^T\boldsymbol{K}\boldsymbol{f} &=\boldsymbol{K}^T\boldsymbol{g}, \notag \\
      \text{cond}(\boldsymbol{K}^T\boldsymbol{K})_2&=\frac{\sigma^2_{\text{max}}}{\sigma^2_{\text{min}}}.
   \end{align}
   \item 
   Regularized inverse problem(set $\boldsymbol{\Gamma}=\boldsymbol{I}$):
   \begin{align}
       (\boldsymbol{K}^T\boldsymbol{K}+\alpha \boldsymbol{I})\boldsymbol{f} &=\boldsymbol{K}^T\boldsymbol{g}, \notag \\
       \text{cond}(\boldsymbol{K}^T\boldsymbol{K}+\alpha \boldsymbol{I})_2&=\frac{\sigma^2_{\text{max}}+\alpha}{\sigma^2_{\text{min}}+\alpha}<\frac{\sigma^2_{\text{max}}}{\sigma^2_{\text{min}}}.
   \end{align}
\end{itemize}

In practice, the regularization parameter $\alpha$ is typically determined by 
identifying the corner of the L-curve, which is defined as the point of maximum 
curvature in the plot of $\log \lVert \boldsymbol{K}\boldsymbol{f} - \boldsymbol{g} \rVert_2^2$ versus $\log \lVert \boldsymbol{\Gamma}\boldsymbol{f} \rVert_2^2$ ~\cite{hansen1992analysis,hansen1993use,engl1988posteriori}. This criterion ensures a balance between the residual and regularization terms, which in physical terms corresponds to mitigating overfitting to noisy data while simultaneously controlling the smoothing imposed by the regularization.~\cite{doi:10.1137/0914086}

Comparing with \cite{Xiong:2025obq}, in this work, we further take into account the error correlations across different values of $\lambda$ in the Tikhonov functional:
\begin{align}
   \boldsymbol{f}_{\text{reg}}^{\alpha}&=
   {\text{arg}}\min_{\boldsymbol{{f}}} [(\boldsymbol{K}\boldsymbol{{f}} - {\boldsymbol{g}})^{T} \boldsymbol{C}_{\boldsymbol{{g}}}^{-1}(\boldsymbol{K}\boldsymbol{{f}} - {\boldsymbol{g}}) + \alpha||\boldsymbol{{f}}||^2_2], \notag \\
   &=(\boldsymbol{K}^{T} \boldsymbol{C}_{\boldsymbol{{g}}}^{-1}\boldsymbol{K}+\alpha \boldsymbol{I})^{-1}\boldsymbol{K}^{T}\boldsymbol{C}_{\boldsymbol{{g}}}^{-1}{\boldsymbol{g}}, \label{TR_reg}
\end{align}
where $\boldsymbol{C}_{{\boldsymbol{g}}}$ is the covariance matrix of different $\lambda$ of a data with $N$ samples.

\subsection{The Backus-Gilbert method}\label{BGmethod}
In 1968, G.Backus and F.Gilbert proposed an approach to tackle linear moment problems~\cite{backus-gilbert}. Their method, now known as the Backus-Gilbert (BG) method, was first applied in geology for mass density estimation and was subsequently extended to a wide range of inverse problems in physics and engineering, such as the study of spectral function in lattice QCD, scatterometer backscatter imaging in geophysics~\cite{PhysRevD.101.114503,backscatter}.

For a linear inverse problem of the form:
\begin{align}
   {g}_i \equiv g(\lambda_i) 
   = \int_{x_{\text{min}}}^{x_{\text{max}}} dx \, K(x,\lambda_i) f_{\text{true}}(x),
\end{align}
which describe the L-IDFT we discussed. The BG method yields an estimated solution $f_{\text{est}}(x')$, which is expressed as a linear combination of the data:
\begin{align}
   f_{\text{est}}(x')&=\sum_{i}a_{i}(x'){g}_i, \notag  \\
   &=\sum_{i}\int_{x_{\text{min}}}^{x_{\text{max}}}dxa_{i}(x')K(x,\lambda_i)f_{\text{true}}(x),\notag \\ 
   &=\int_{x_{\text{min}}}^{x_{\text{max}}}dx \rho(x-x')f_{\text{true}}(x). \label{eq_resolution}
\end{align}
The $a_i(x)$ denotes the coefficients to be determined, and $\rho(x-x')=\sum_{i}a_{i}(x')K(x,\lambda_i)$ is named as averaging kernel function~\cite{ASTER2019135} or resolution function~\cite{Karpie2019}, which satisfies  the normalized condition: $1=\int_{x_{\text{min}}}^{x_{\text{max}}}dx \rho(x-x')$. Eq.~(\ref{eq_resolution}) shows that as $\rho(x-x')\rightarrow \delta(x-x')$, the estimated solution $f_{\text{est}}(x)$ will approach to the true solution $f_{\text{true}}(x)$. In other words, a narrower width $l$ of $\rho(x-x')$ yields a better approximation to $f_{\text{true}}$. Building on these ideas, the BG method seeks an optimal solution by minimizing the width $l$, which is defined as:
\begin{align}
   l(x')&=\int_{x_{\text{min}}}^{x_{\text{max}}}dx(x-x')^2 (\rho(x-x'))^2.
\end{align}
For convenience in identifying the minimal width, one can rewrite the above expressions as:
\begin{align}
   {a}_i&=a_i(x'), \notag\\
   {m}_i&=\int_{x_{\text{min}}}^{x_{\text{max}}}dx K(x,\lambda_i), \notag\\
   {H}_{ij}(x')&=\int_{x_{\text{min}}}^{x_{\text{max}}}dx(x-x')^2K(x,\lambda_i)K(x,\lambda_j). 
\end{align}
Therefore, the normalization condition is $\boldsymbol{a}^T \boldsymbol{m} = 1$, 
with the width $l(x') = \boldsymbol{a}^T \boldsymbol{H} \boldsymbol{a}$ and the 
estimated solution $f_{\text{est}}(x') = \boldsymbol{a}^T \boldsymbol{g}$. 
The problem then reduces to finding the optimal vector $\boldsymbol{a}$ that minimizes the width, subject to the normalization condition:
\begin{align}
   \boldsymbol{a}_{\text{op}}=\text{arg}\min_{\boldsymbol{a}} [\boldsymbol{a}^T\boldsymbol{H}\boldsymbol{a}+\beta (\boldsymbol{a}^T\boldsymbol{m}-1)].
\end{align}
Similar to the derivation of the regularized solution in Tikhonov regularization, we have:
\begin{align}
   \boldsymbol{a}_{\text{op}}=-\frac{1}{2}\beta \boldsymbol{H}^{-1}\boldsymbol{m} =\frac{1}{\boldsymbol{m}^T \boldsymbol{H}^{-1}\boldsymbol{m}}\boldsymbol{H}^{-1}\boldsymbol{m}.
\end{align}
Here, we have used $\beta=-2\left [\boldsymbol{m}^T\boldsymbol{H}^{-1}\boldsymbol{m}\right]^{-1}$, which is determined by the normalization condition $\boldsymbol{a}_{\text{op}}^T\boldsymbol{m}=1$.

In the original formulation by Backus and Gilbert, the ill-conditioning of the matrix $H$ was not explicitly addressed. In practice, however, this issue must be mitigated through additional regularization techniques. Accordingly, the modified functional, together with its corresponding optimal vector $\boldsymbol{a}$, is defined as:
\begin{align}
L[\boldsymbol{a}]&=\boldsymbol{a}^T\boldsymbol{H}\boldsymbol{a}+\alpha\boldsymbol{a}^T\boldsymbol{C_{{g}}}\boldsymbol{a}+\beta (\boldsymbol{a}^T\boldsymbol{m}-1), \notag\\
   \boldsymbol{a}_{\text{op}}&=\frac{1}{\boldsymbol{m}^T (\boldsymbol{H}+\alpha\boldsymbol{C_{{g}}})^{-1}\boldsymbol{m}}(\boldsymbol{H}+\alpha\boldsymbol{C_{{g}}})^{-1}\boldsymbol{m}.\label{Bg}
\end{align}
Here, the term $\boldsymbol{a}^T \boldsymbol{C}_{g} \boldsymbol{a}$ serves as a regularization penalty that stabilizes the estimated solution $f_{\text{est}}(x')$. The parameter $\alpha$ thus acts as a trade-off factor, balancing stability against bias in the reconstructed solution. The philosophy underlying this approach is analogous to Tikhonov regularization. The key difference, lies in the mathematical form employed to quantify the bias: Tikhonov regularization measures the deviation through $\lVert \boldsymbol{K}\boldsymbol{f} - \boldsymbol{g} \rVert_2^2$, whereas the Backus–Gilbert method uses $\boldsymbol{a}^T \boldsymbol{H} \boldsymbol{a}$.

The Backus-Gilbert method provides an estimate of the solution on a grid of points $\left \{ x_j \right \}_{j=0}^m $. However, since the averaging kernel $\rho(x-x_j)$ may not be well approximated to $\delta(x-x_j)$ (primarily due to the limited number of data points provided by current lattice simulation). Such a contamination of averaging kernel may vary and overlap in complicated ways~\cite{ASTER2019135}. Therefore the Backus–Gilbert method may yield unsatisfactory results in L-IDFT.

\subsection{Bayesian approach with GRW}\label{Bayesianapproach}
Following the introduction of classical regularization approaches to inverse problems, we next consider the Bayesian framework. In this formulation, the unknown solution is modeled as a random variable characterized by a probability distribution, whose effectiveness has been widely demonstrated in practical applications~\cite{PhysRevLett.111.182003,Bayesian_Inverse_problem_application,10.7551/mitpress/7132.001.0001}. 
More recently, a Gaussian Process Regression (GPR) method based on the Bayesian approach has been applied to the same problem using PDF data, and it successfully reconstructed the solution in momentum space~\cite{Medrano:2025cmg}.
The Bayesian perspective offers a fresh conceptual lens through which to view many classical regularization techniques, since the regularization term can be naturally encoded in the choice of prior distribution~\cite{Stuart_2010}. The foundation of the Bayesian framework is Bayes' theorem, which can be written as:
\begin{align}
   P(H_i|E)&=\frac{P(E|H_i)\cdot P(H_i)}{P(E)}, \notag\\
   &=\frac{P(E|H_i)\cdot P(H_i)}{\sum_{i}P(E|H_i)\cdot P(H_i)}, \label{eq_bayesiantheorem} 
\end{align}
where $H_i$ denotes a hypothesis and $E$ denotes the observed evidence. 
Here, $P(E|H_i)$ is the likelihood, quantifying the probability of observing the evidence $E$ given that $H_i$ is true. 
$P(H_i)$ is the prior probability, representing the plausibility of $H_i$ before considering the current evidence. $P(H_i|E)$ is the posterior probability, which provides the updated probability of $H_i$ being true after taking the evidence $E$ into account.
Finally, $P(E)$ is the marginal probability of the evidence $E$, serving as a normalization constant that ensures the posterior probability $P(H_i|E)$ forms a normalized probability distribution. It is computed by applying the formula of total probability to the numerator, as shown in Eq.~(\ref{eq_bayesiantheorem}).

In the Bayesian framework for inverse problems, the candidate solution $\boldsymbol{f}$ in the solution space is treated as the hypothesis $H_i$ to be inferred, while the the measured data $\boldsymbol{g}$ is regarded as the observed evidence $E$. As a result, the Bayesian approach takes the form:
\begin{align}
p(\boldsymbol{f}|\boldsymbol{g},I)&=\frac{p(\boldsymbol{g}|\boldsymbol{f},I)\cdot p(\boldsymbol{f}|I)}{\int_{\text{possible solutions}}p(\boldsymbol{g}|\boldsymbol{f},I)\cdot p(\boldsymbol{f}|I)d\boldsymbol{f}}, \notag \\
&\propto p(\boldsymbol{g}|\boldsymbol{f},I)\cdot p(\boldsymbol{f}|I).
\end{align}
Since $\boldsymbol{f}$ is a continuous random variable, the notation $p(\cdot)$ denotes a probability density function, in contrast to the probability mass function $P(\cdot)$ used for discrete random variables. Consequently, the summation in the denominator in Eq.~(\ref{eq_bayesiantheorem}) is replaced by an integral over the continuous domain of $\boldsymbol{f}$.
The $I$ denotes the background information or assumptions under which the inference is performed. The symbol $I$ is not a random variable but rather a shorthand for the set of conditions, constraints, and theoretical or practical knowledge that are assumed to hold. 
Such background information may include physical principles, structural assumptions, or the statistical model for measurement errors~\cite{Bayesian_Inverse_problem_application}. For instance, given that the $x$-space quasi-DA is theoretically expected to be smooth~\cite{LatticeParton:2022zqc}, a smoothness prior $p(\boldsymbol{f}|I)$ can be imposed in the reconstruction of the quasi-DA. Writing the prior explicitly as $p(\boldsymbol{f}|I)$ highlights that the distribution is not arbitrary, but conditioned on the specific modeling context.

We now present the mathematical formulation of likelihood and prior distribution. For the sample data ${\scriptsize \left \{\boldsymbol{g}^{(h)} \right \} }_{h=1}^N$ with zero-mean Gaussian noise $\boldsymbol{\bar{\delta}}$:
   \begin{align}
       \bar{\boldsymbol{\delta}}&=\bar{\boldsymbol{g}}-\boldsymbol{g}_{\text{true}}, \notag \\
       &=\bar{\boldsymbol{g}}-\boldsymbol{K\boldsymbol{f}_{\text{true}}}
       \sim \mathcal{N}(\boldsymbol{0}, \boldsymbol{C}_{\boldsymbol{\bar{g}}}),
   \end{align}
where
   \begin{align}
       \boldsymbol{C}_{\boldsymbol{\bar{g}}}=\frac{1}{N}\boldsymbol{C}_{\boldsymbol{g}}.
   \end{align}
Consequently, the distribution of the sample mean conditional on the true solution is:
   \begin{align}
       p(\bar{\boldsymbol{g}}|\boldsymbol{f}_{\text{true}},I)\propto\text{exp}\small\left [-\frac{1}{2}(\boldsymbol{\bar{g}}-\boldsymbol{K}\boldsymbol{f_{\text{true}}})^T \boldsymbol{C}_{\boldsymbol{
       \bar{g}}}^{-1}(\boldsymbol{\bar{g}}-\boldsymbol{K}\boldsymbol{f_{\text{true}}})\ \right].
   \end{align}
In statistical inference, however, $\boldsymbol{f}_{\text{true}}$ is unknown and treated as a fixed but inaccessible quantity. To evaluate how plausible different candidate solutions of $\boldsymbol{f}$ are given the sample mean $\boldsymbol{\bar{g}}$, we define the likelihood distribution by replacing $\boldsymbol{f}_{\text{true}}$ with $\boldsymbol{f}$:
   \begin{align}
    p(\bar{\boldsymbol{g}}|\boldsymbol{f},I)\propto\text{exp}\small\left [-\frac{1}{2}(\boldsymbol{\bar{g}}-\boldsymbol{K}\boldsymbol{f})^T \boldsymbol{C}_{\boldsymbol{
       \bar{g}}}^{-1}(\boldsymbol{\bar{g}}-\boldsymbol{K}\boldsymbol{f})\ \right].\label{bayesian_likeli}
   \end{align}

Another important part is the formulation of the prior distribution $p(\boldsymbol{f}|I)$. For problems in this work, solutions possess theoretical smoothness. Therefore, we adopt a Gaussian Random Walk (GRW) prior to characterize this smoothness property~\cite{Lawler_Limic_2010}, whose mathematical formulation is as follows:
   \begin{align}
       p(\boldsymbol{f}|I)&=\int p(\boldsymbol{f}|\sigma,I)\cdot p(\sigma|I)d\sigma, \notag\\
       p(\boldsymbol{f}|\sigma,I)&=p(f_0|I)\cdot \prod_{j=1}^{m}\frac{1}{\sqrt{2\pi\sigma^2}}e^{-\frac{(f_j-f_{j-1})^2}{2\sigma^2}}, \notag\\
       p(\sigma|I)&=
      \begin{cases}
         \frac{2}{\sqrt{2\pi\eta^2}}e^{-\frac{\sigma^2}{2\eta^2}}, \quad &\text{if} \quad \sigma>0,
         \\0,\quad &\text{otherwise}.
      \end{cases}\label{eq_Bayesprior}
   \end{align}
Here $f_j$ represents $f(x_j)$, and $p(f_0|I)$ characterizes the prior knowledge about the initial value $f(x_0)$ of the random walk. The hyperparameter $\sigma$ governs the step size of the random walk between adjacent points. In this Bayesian framework, $\sigma$ is analogous to the regularization parameter in conventional regularization approaches, as both control the trade-off between data fitness and the smoothness constraint imposed on the solution; with smaller values of $\sigma$ corresponding to a stronger smoothness constraint on the solution. Nevertheless, while they serve a similar purpose, there is no unambiguous criterion for uniquely determining the value of hyperparameter within the Bayesian framework. Consequently, we adopt a hierarchical Bayesian model~\cite{Gelman2013} by treating the hyperparameter as an unknown random variable and assigning it a probability distribution $p(\sigma|I)$ of a specific form, as shown in Eq.~(\ref{eq_Bayesprior}), where $\eta$ is the scale parameter. The $p(\sigma|I)$ is commonly known as hyper-prior.

Having defined the likelihood and prior distributions, we ultimately obtain the posterior distribution, which contains all available information about the solution to the inverse problem. 
To generate samples from the posterior distribution for inference, we employ Markov Chain Monte Carlo (MCMC) methods~\cite{Stuart_2010, ASTER2019135}. The posterior mean estimation~\cite{Bayesian_Inverse_problem_application} is inherently compatible with MCMC sampling, as the algorithm naturally converges to the posterior distribution, making it a principled and stable choice for our estimation framework. The solution given by the posterior mean estimation therefore can be defined as:
   \begin{align}
       \boldsymbol{f}_{\text{Bayes}}=\int \boldsymbol{f}\cdot p(\boldsymbol{f}|\boldsymbol{\bar{g}},I)d\boldsymbol{f}.\label{bayesian_solution}
   \end{align}

\begin{widetext}
\begin{figure}[htbp]
    \centering
    \begin{minipage}{2.0\linewidth}
        \centering
        \includegraphics[width=\linewidth]{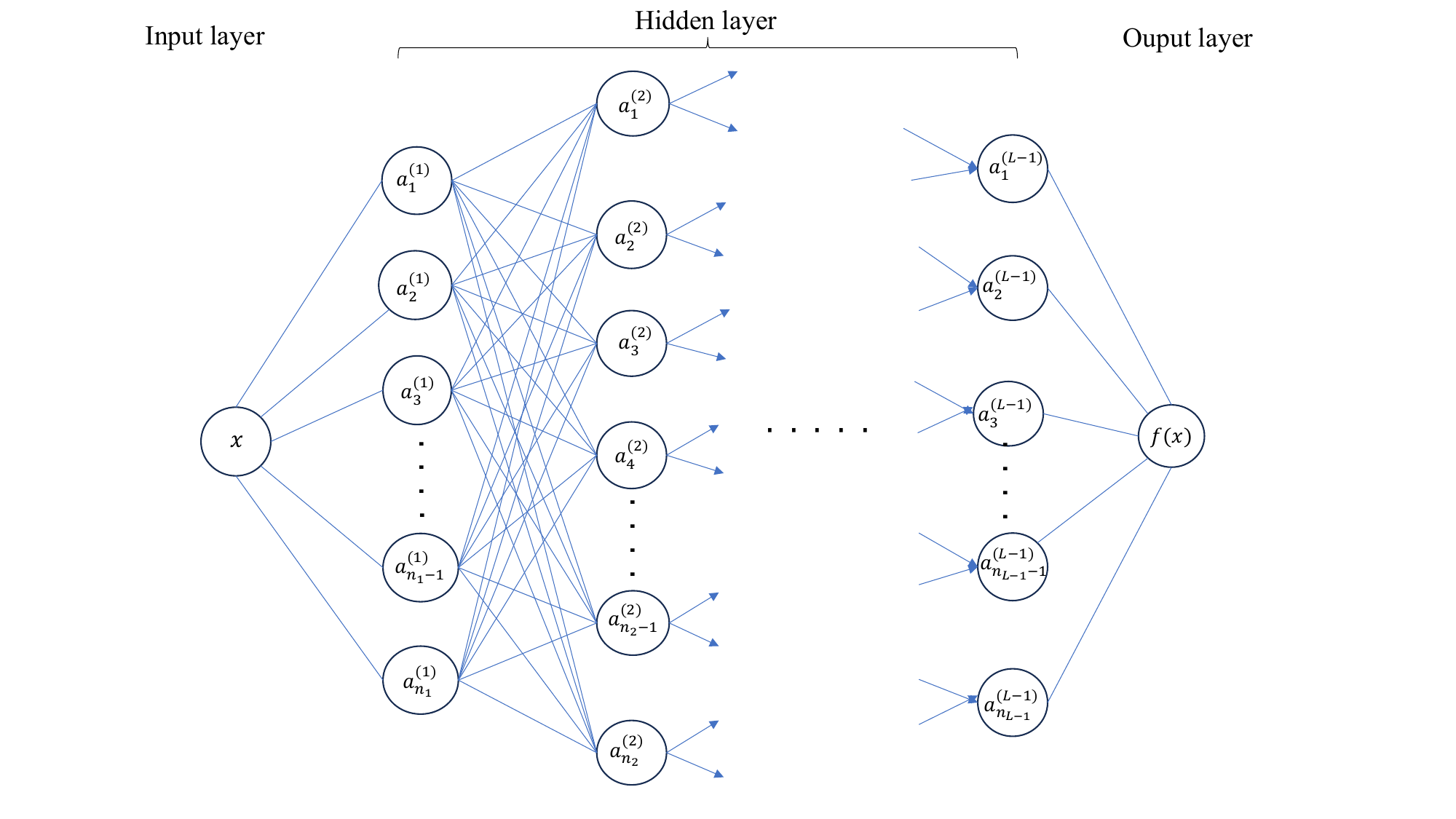}
        \caption{A multilayer feedforward neural network architecture.}
        \label{fig_ANN_picture}
    \end{minipage}
\end{figure}
\end{widetext}
\subsection{Artificial Neuron Networks}
Artificial neural networks (ANNs) have been employed to address complex inverse problems since their inception, and recent advances particularly in the field of imaging have demonstrated their powerful capability in solving such problems. The principal strength of ANNs lies in their expressive non-linear representation and strong generalization capacity, which enables them to approximate highly complex mappings. 
This property renders ANNs particularly well-suited for tackling ill-posed inverse problems, especially in situations where the target functions to be reconstructed lack explicit physical models. 
Consequently, an increasing volume of research has been devoted to applying ANNs in the resolution of diverse inverse problems ~\cite{2023arXiv231014459G, ZHANG2025108544,Chouzenoux_2025}.

The fundamental architecture of a neural network comprises multiple layers of interconnected computational units, referred to as neurons. 
Each neuron receives input signals from neurons in the preceding layer, performs a weighted aggregation of these inputs, and applies a nonlinear transformation to generate its own output, typically implemented via an activation function. The connection weights and bias terms within each neuron constitute the model's adjustable parameters. These parameters are optimized iteratively through training algorithms that leverage data-driven feedback, enabling the network to progressively acquire the capacity to extract relevant features from inputs 
and to approximate complex functional mappings. In this study, we employ a multilayer feedforward neural network (FNN) to address the inverse problem, which has been recently applied for researches for PDFs and GPDs~\cite{Chu:2025jsi}.

The FNN is a deep learning model based on a hierarchical architecture, characterized by a unidirectional flow of information that forms a directed acyclic graph. The network as shown in Fig.~\ref{fig_ANN_picture} comprises an input layer, multiple hidden layers and an output layer, 
with each layer containing a set of neurons. Neurons in adjacent layers are fully connected, while there are no connections between neurons in the same layer. Each neuron computes a linearly weighted summation of its inputs, adds a bias term, and subsequently applies a nonlinear activation function (such as ReLU or Sigmoid) to produce its output. 
This operation can be formally expressed as:
\begin{align}
    a_{j}^{(L)} = \sigma^{(L)}\left( \sum_{i=1}^{N_{L-1}} w_{ji}^{(L)} a_i^{(L-1)} + b_j^{(L)} \right),
\end{align}
where $a_j^{(L)}$ denotes the activation of the $j$-th neuron in the $L$-th layer, $w_{ji}^{(L)}$ represents the synaptic weight connecting the $i$-th neuron in layer $L-1$ to the $j$-th neuron in layer $L$, and $b_j^{(L)}$ is the bias term associated with that neuron. The term $\sum_{i=1}^{N_{L-1}} w_{ji}^{(L)} a_i^{(L-1)}$ represents the weighted sum of inputs from the previous layer. Adding the bias and applying the activation $\sigma^{(L)}(\cdot)$ introduce nonlinearity, enabling the network’s expressive capability.

Accordingly, a feedforward neural network with $L$ layers defines its overall mapping 
$f$ from input $x$ to output $f(x)$ as a composition of transformations:
\begin{align}
    f(x) = (\sigma^{(L)} \circ A^{(L)} \circ \cdots \circ \sigma^{(2)} \circ A^{(2)} 
           \circ \sigma^{(1)} \circ A^{(1)})(x), \label{eq:NNmap}
\end{align}
where $A^{(L)}$ denotes the affine transformation 
$\left( \sum_{i=1}^{N_{L-1}} w_{ji}^{(L)} a_i^{(L-1)} + b_j^{(L)} \right)$ at the $L$-th layer, and $\circ$ represents the composition of mappings. Thus, the function $f(x)$ is implicitly determined by the set of parameters $\{w\}$ and $\{b\}$ in the feedforward neural network.

To solve the inverse problem in this work, one can employ a genetic algorithm or gradient descent algorithm
to search the parameter space of weights $\{w\}$ and biases $\{b\}$, 
with the objective of identifying the configuration that minimizes 
the loss function defined as:
\begin{align}
  \chi^2(\{w\},\{b\}) 
  &= \left[ 
        (\boldsymbol{K}\boldsymbol{f}_{\{w\},\{b\}} - \boldsymbol{g})^T 
        C_{\boldsymbol{g}}^{-1} 
        (\boldsymbol{K}\boldsymbol{f}_{\{w\},\{b\}} - \boldsymbol{g}) 
     \right].\label{eq_ANN}
\end{align}
The $C_{\boldsymbol{g}}$ denotes the covariance matrix of the observed data $\boldsymbol{g}$. The minimization of the loss function $\chi^2(\{w\},\{b\})$ yields the optimal configuration of the network parameters. 
This configuration defines the reconstructed function $\boldsymbol{f}_{\{w\},\{b\}}$, which best represents the target function associated with the input data $\boldsymbol{g}$.

\section{Numerical results}\label{sec_numresultes}
In this section, we present a comprehensive numerical study to evaluate the performance of the four reconstruction approaches considered in this work. 
We begin by validating their fundamental functionality and stability using two controlled toy models with known ground truths. 
The first toy model employs the asymptotic form of the distribution amplitude as the reference solution. 
This choice is motivated by its close connection to the physical problem of interest, allowing for a direct assessment of the accuracy and stability of the reconstruction approaches under conditions that closely resemble realistic applications. 
In contrast, the second toy model adopts a more intricate, non-standard functional form as the ground truth. 
This setup is designed to probe the generality and robustness of the methods when confronted with a more challenging reconstruction task, serving as a preliminary step toward future investigations of more complex distribution amplitudes. Finally, the four approaches are applied to actual lattice quasi distribution of LCDA to examine their practical effectiveness and to compare the reconstructed results in a physically relevant setting.

\subsection{Toy model \uppercase\expandafter{\romannumeral1}}
To align with the problem we intend to address in practice, we adopt a specific test function based on the asymptotic form, which is defined as follows and serves as the true solution in our analysis:
\begin{align}
  f(x)=
  \begin{cases}
  \frac{\Gamma(0.5+1+2)}{\Gamma(0.5+1)\Gamma(1+1)}x^{0.5}(1-x), &  x\in [0,1] \\
   0, &\text{otherwise} 
  \end{cases}.\label{truesolution1}
\end{align}
In which these $\Gamma$ functions are normalized factor. We present the numerical example with the following parameters:
  \begin{align}
      &x_{\text{min}}=0,\quad x_{\text{max}}=1,\quad \Delta x=0.01, \notag\\
      &\lambda_{\text{min}}=-15,\quad \lambda_{\text{max}}=15, \quad \Delta\lambda=0.5.
  \end{align}
The distribution in coordinate space $g(\lambda)$ with errors are constructed as:
\begin{align}
  g^{\delta}(\lambda_k)&=g(\lambda_k)+\delta_{\text{uncorr}}(\lambda_k)+\delta_{\text{corr}}(\lambda_k).
\end{align}
Here, the distribution $g(\lambda_k)$ is obtained using the discrete Fourier transform (DFT) defined in Eq.~(\ref{eq_mapFT}), while $\delta_{\text{uncorr}}(\lambda_k)$ and $\delta_{\text{corr}}(\lambda_k)$ represent uncorrelated errors and correlated errors, respectively. The $\delta_{\text{uncorr}}(\lambda_k)$ is given by:
\begin{align}
  \delta_{\text{uncorr}}(\lambda_k)&\sim e^{M_{\text{uncorr}}|\lambda_k|}\cdot\mathcal{N}(\mu=0,\sigma_{\text{uncorr}}^2),
\end{align}
where $\mathcal{N}(\mu=0, \sigma_{\text{uncorr}}^2)$ stands for a Gaussian distribution with mean $\mu=0$ and standard deviation $\sigma_{\text{uncorr}}=0.1$. In addition, an exponential increasing factor $e^{M_{\text{uncorr}}|\lambda_k|}$ with $M_{\text{uncorr}}=0.15$ is introduced to characterize the noise increasing behavior at large Euclidean non-local separations, analogous to hadronic matrix element measurements in lattice QCD. 
On the other hand, the correlated errors is modeled within a Gaussian process (GP) framework, which provides a natural way to describe spatially correlated fluctuations in lattice data. In this approach, the correlated error term $\delta_{\text{corr}}(\lambda)$ is treated as a stochastic Gaussian function $\mathcal{GP}(\cdot,\cdot)$ characterized by the covariance
\begin{equation}
    \delta_{\text{corr}} \sim \mathcal{GP}(0, k(\lambda_i, \lambda_j)),
\end{equation}
where $k(\lambda_i, \lambda_j)$ specifies how errors at different points $\lambda_i$ and $\lambda_j$ are correlated. The covariance kernel is defined as
{\small
\begin{equation}
    k(\lambda_i, \lambda_j)
    = \sigma_{\text{corr}}^2
    \exp\!\Big[-\frac{(\lambda_i - \lambda_j)^2}{2\ell^2}\Big]
    \exp\!\Big[M_{\text{corr}}(|\lambda_i| + |\lambda_j|)\Big],
\end{equation}
}with $\sigma_{\text{corr}}^2 = 4\times10^{-4}$ setting the overall variance scale, $\ell = 3$ controlling the correlation length, and $M_{\text{corr}} = 0.1$ describing the gradual increase of the correlated uncertainty at large Euclidean separations. The first exponential factor $\exp\!\Big[-\frac{(\lambda_i - \lambda_j)^2}{2\ell^2}\Big]$ encodes the spatial correlation, ensuring that nearby points have similar fluctuations over the characteristic scale $\ell$. The second exponential $\exp[M_{\text{corr}}(|\lambda_i| + |\lambda_j|)]$ introduces a smooth amplification of the correlated error magnitude at large $|\lambda|$, accounting for the increased uncertainty in that region.

\begin{figure}
    \centering
    \includegraphics[width=0.8\linewidth]{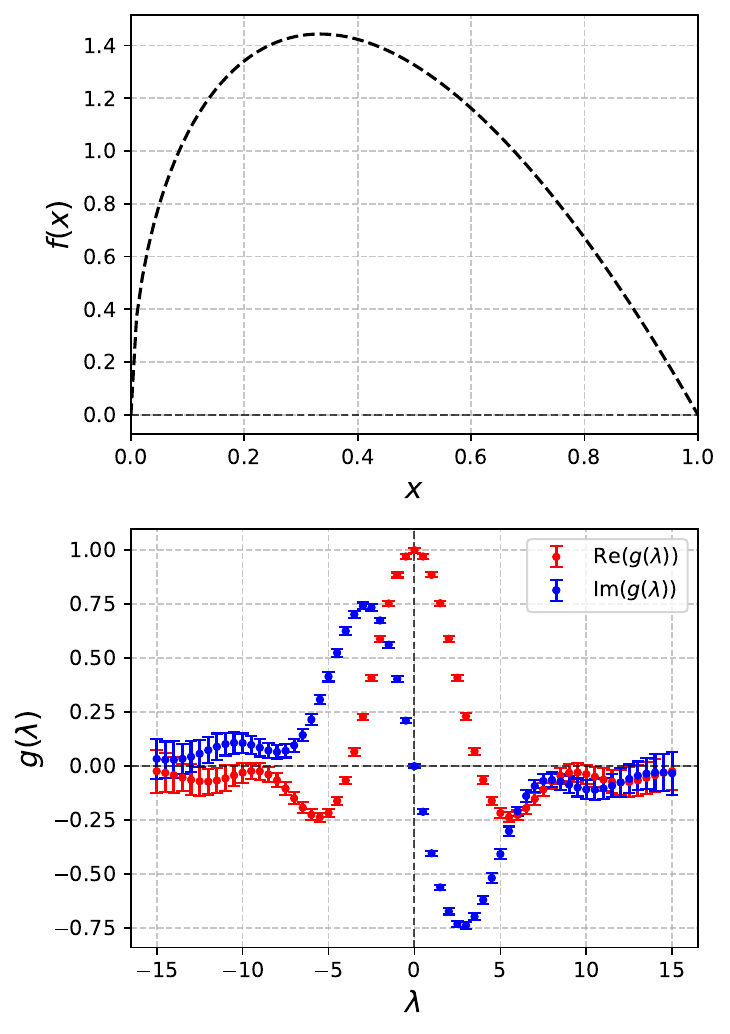}
    \caption{Upper panel: The true solution is taken to be the asymptotic form of the distribution amplitude. Lower panel: The error-contaminated distribution $g^{\delta}(\lambda)$ in coordinate space.}
    \label{fig_fake_data1}
\end{figure}

\begin{figure}
    \centering
    \includegraphics[width=0.9\linewidth]{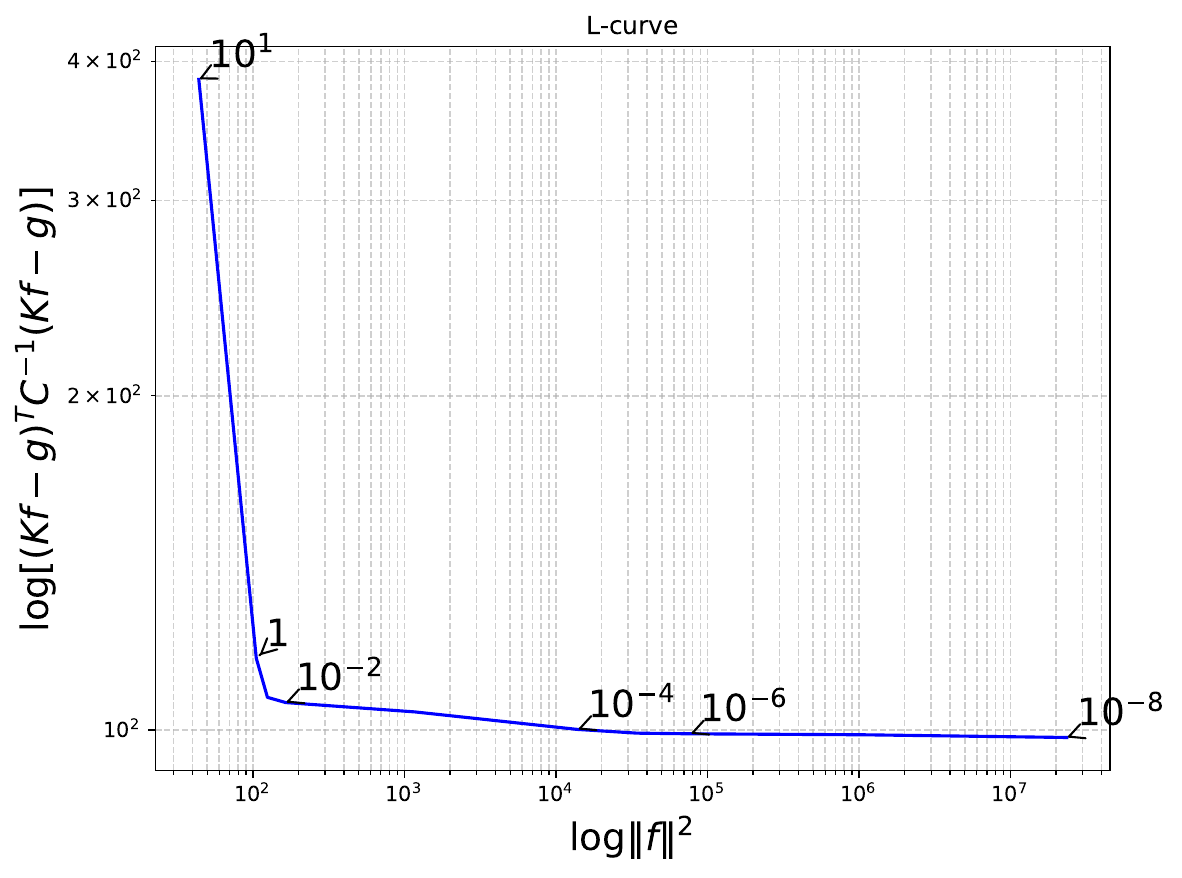}
    \caption{L-curve for determining the regularization parameter $\alpha$ in Tikhonov method for Toy model {\uppercase\expandafter{\romannumeral1}}}
    \label{fig_L_curve_1_TR}
\end{figure}

Therefore, Fig.~\ref{fig_fake_data1} illustrates the assumed true solution of our toy model in momentum space, together with the corresponding coordinate-space data obtained via a discrete Fourier transform and subsequently contaminated with artificial statistical noise. Based on this noisy dataset, we apply the four reconstruction methods discussed above to solve the L-IDFT inverse problem. For the Tikhonov regularization, Backus–Gilbert method, Bayesian approach with GRW prior and feedforward ANN, we implement the formulations given in Eq.~(\ref{TR_reg}), Eq.~(\ref{Bg}), Eqs.~(\ref{bayesian_likeli}) to (\ref{bayesian_solution}), and Eq.~(\ref{eq_ANN}) respectively. The respective parameters are set as follows: 
\begin{itemize}
    \item Tikhonov: $\alpha_{\text{TR}}=0.1$(determined from the L-curve in Fig.~\ref{fig_L_curve_1_TR}).
    \item Backus-Gilbert: $\alpha_{\text{BG}}=0.05$.
    \item Bayesian: \\
    (1) Gaussian distribution $\mathcal{N}(0,0.1^2)$ for $p(f_0|I)$ (based on the physical expectation of vanishing endpoint values).\label{initprior}\\
    (2) Hyper-prior scale parameter $\eta$=0.01 (see Appendix.~\ref{sec_hyperprior_determination} for details on the parameter selection).
    \item ANN: Single hidden layer of 5 neurons with the Sigmoid activation function $\text{Sigmoid}(t)=\frac{1}{1+e^{-t}}$, employing a genetic algorithm for parameter optimization.
\end{itemize}

\begin{figure}
    \centering
    \includegraphics[width=1.0\linewidth]{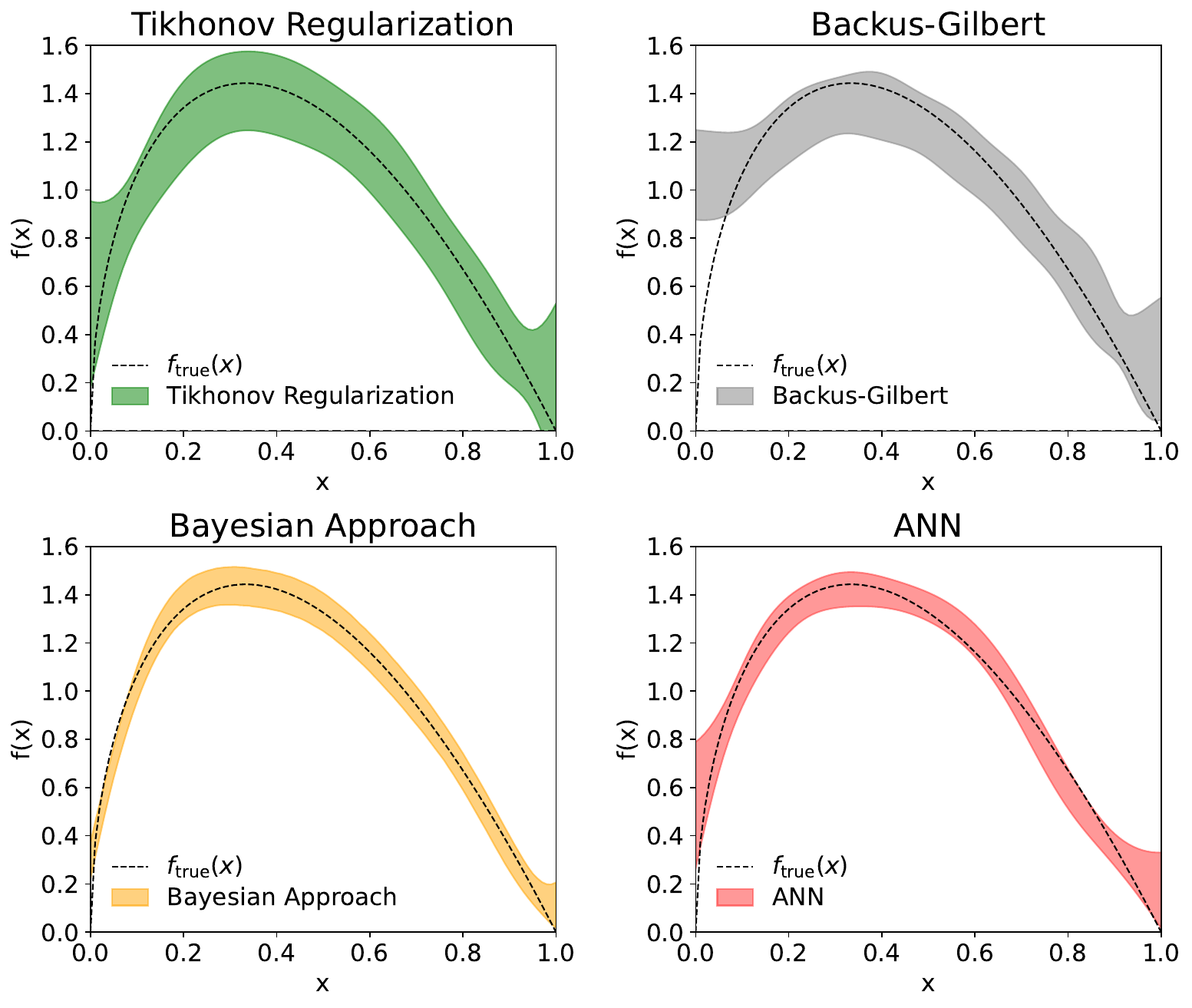}
    \caption{Reconstruction results of Toy model {\uppercase\expandafter{\romannumeral1}} obtained by four methods.}
    \label{fig_Toymodel1_result}
\end{figure}
\begin{figure}
    \centering
    \includegraphics[width=0.8\linewidth]{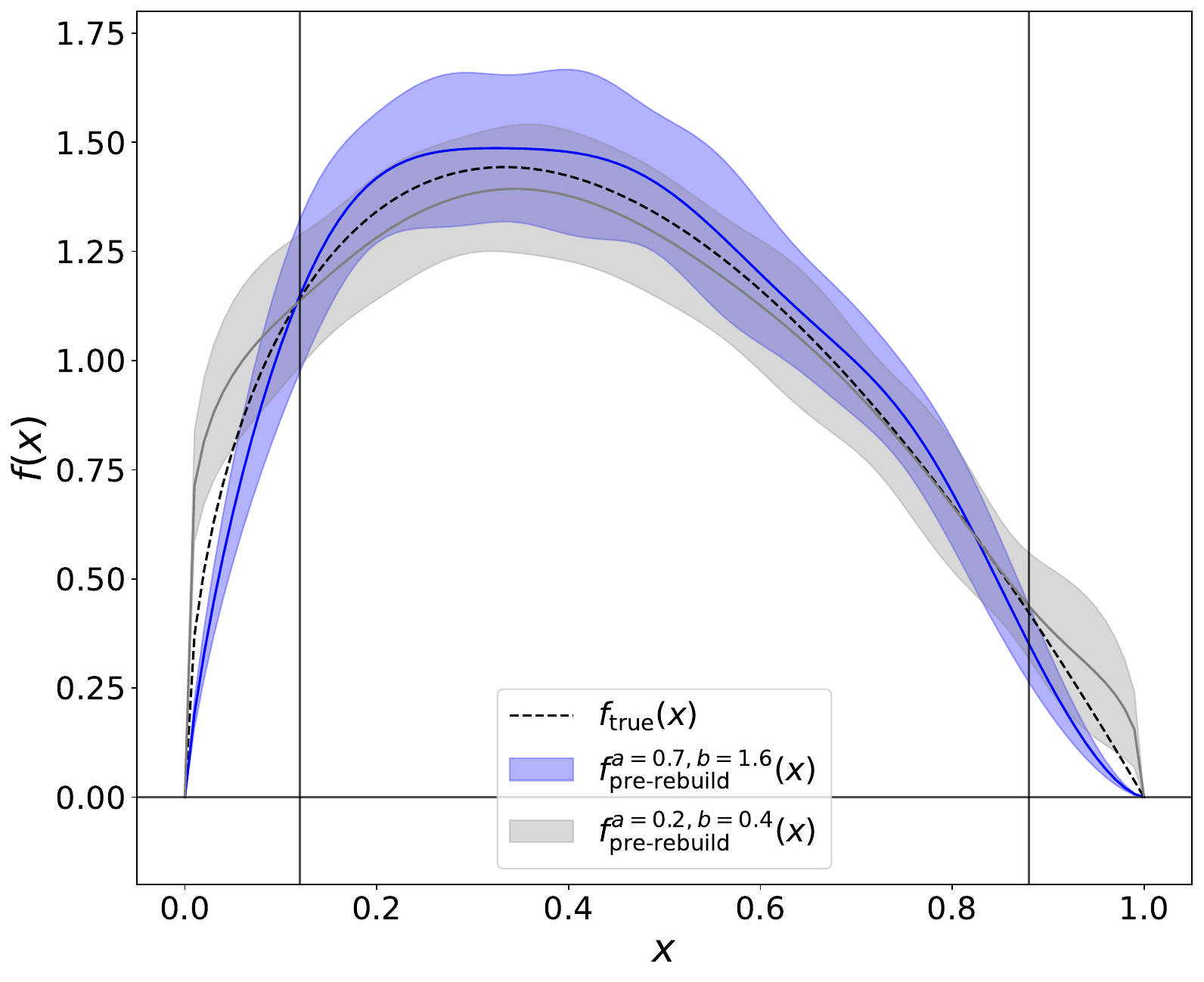}
    \caption{The reconstruction obtained by precondition-BG using different precondition function. Blue band: $a=0.7,b=1.6$. Gray band: $a=0.2,b=0.4$. Blue line: mean value of $f_{\text{pre-rebuild}}^{a=0.7,b=1.6}(x)$. Gray line: mean value of $f_{\text{pre-rebuild}}^{a=0.2,b=0.4}(x)$. Dash line: true solution.}
    \label{fig_pre_BG}
\end{figure}

As shown in the Fig.~\ref{fig_Toymodel1_result}, the result obtained by BG only manages to roughly capture the behavior of the true solution within the region $x\in[0.2,1]$, while it exhibits significant deviation from the true solution in the region $x \in [0, 0.2]$, thereby validating the previous expectation of the BG method in section \ref{BGmethod}. In practice, a consequence of insufficient data in the large-$\lambda$ region that prevents the resolution function from sufficiently approximating the $\delta$-function. 
We note that Ref.~\cite{Karpie2019} employed a preconditioning function $f_{p}(x)=Nx^{a}(1-x)^{b}$ with normalized constant $N$ to improve the result within the region $x \in [0, 0.2]$, while the parameters $a, b$ are determined by fitting to the data in coordinate space. However, this modified approach renders that the BG solution is highly sensitive to the specific form of the preconditioning. On the other hand, the parameters $a$ and $b$ obtained by fitting over the entire $\lambda$ range are not physically reasonable. As demonstrated in studies of extrapolation method \cite{Gao:2021dbh}, such asymptotic behavior is only valid in the large $\lambda$ region. To validate this concern, we also conducted a numerical test starting from the coordinate-space data with different preconditions.  As shown in Fig.~\ref{fig_pre_BG}, the BG method exhibits sensitivity to the precondition, and introduces significant systematic uncertainty, especially near the endpoint regions.


In contrast, the other three methods successfully reconstruct the true solution across almost the entire region. In fact, these three methods share a common underlying logic in effectively solving the L-IDFT inverse problem. 
\begin{itemize}
    \item The are all fundamentally based on regularization techniques. In the cases of Tikhonov regularization and the Bayesian approach, the regularization is explicitly introduced through a regulator term. Whereas in the artificial neural network (ANN), the number of neurons and parameters implicitly restricts the degrees of freedom of the solution, playing an equivalent role to regularization. 
    \item They are all based on a data-fidelity function that quantitatively measures the discrepancy between the reconstructed and true (or observed) data. The ANN enforces consistency with the input data and drives the reconstructed solution toward the true one by explicitly defining a loss function. Similarly, both Tikhonov regularization and the Bayesian approach contain the term $(\boldsymbol{K}\boldsymbol{{f}} - {\boldsymbol{g}})^{T} \boldsymbol{C}^{-1}(\boldsymbol{K}\boldsymbol{{f}} - {\boldsymbol{g}})$ in the functional and likelihood formulations respectively, which serves the same purpose of guiding the solution to globally approximate the true distribution.
\end{itemize}

It can be observed that the reconstruction results still exhibit some deviations around the endpoint regions $x < 0.1,  x > 0.9$. These deviations occur because the regularization applied in our methods do not impose strong prior constraints on the solution's behavior near the endpoints. As a result, the inherent ill-posedness in these regions is not adequately regularized, which consequently leads to the observed lack of convergence to the true solution at the boundaries. However, it should be emphasized that these local deviations do not violate the global mathematical convergence of the methods, and the overall reconstruction remains reliable over the majority of the entire distribution.
However, it should be noted that the Bayesian approach yields significantly smaller deviations near the endpoints compared to the other approaches. This improvement can be attributed to the prior constraint explicitly imposed on the initial value $f(x_0)$, as detailed in section~\ref{Bayesianapproach}. This demonstrates the Bayesian approach's capability to incorporate prior knowledge in a flexible manner, thereby enabling a richer and more physically motivated regularization structure than the Tikhonov scheme.

Overall, the uncertainties associated with the Bayesian and ANN reconstructions are comparable and smaller than those obtained with the Tikhonov method. In addition, both approaches produce noticeably smoother results, highlighting their superior stability and regularization efficiency relative to the Tikhonov scheme.

\subsection{Toy model \uppercase\expandafter{\romannumeral2}}
\begin{figure}
    \centering
    \includegraphics[width=0.8\linewidth]{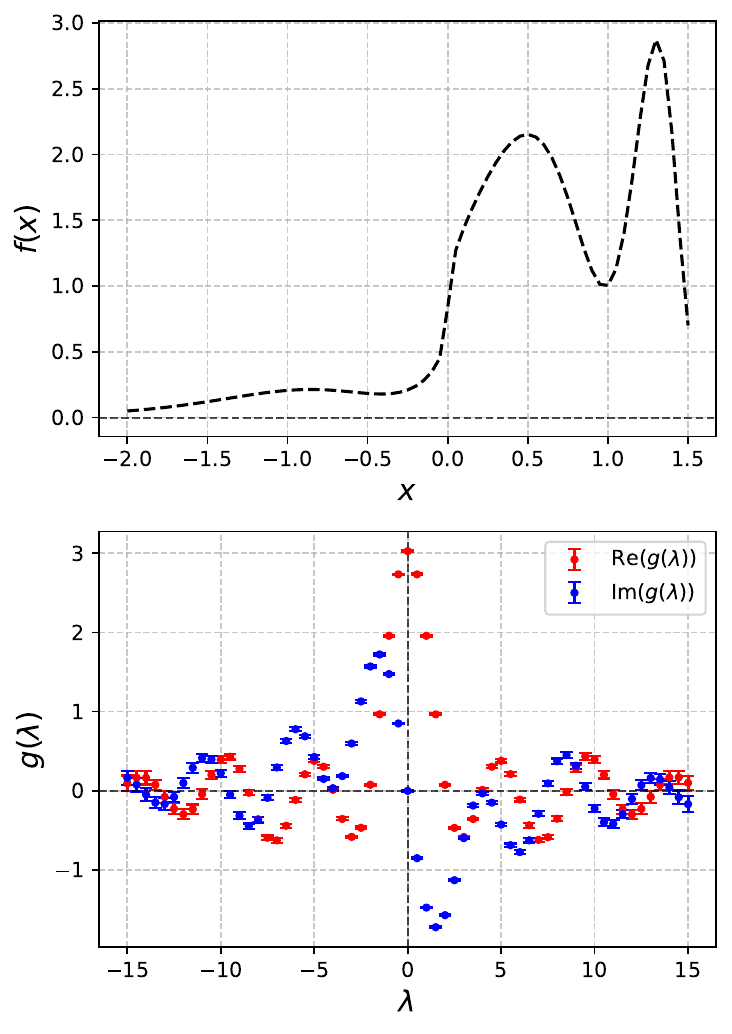}
    \caption{Upper panel: The true solution takes the form of Eq.~(\ref{eq:toymodel2}). Lower panel: The corresponding error-contaminated $g^{\delta}(\lambda)$ in coordinate space.}
    \label{fig_fake_data2}
\end{figure}
\begin{figure}
    \centering
    \includegraphics[width=0.9\linewidth]{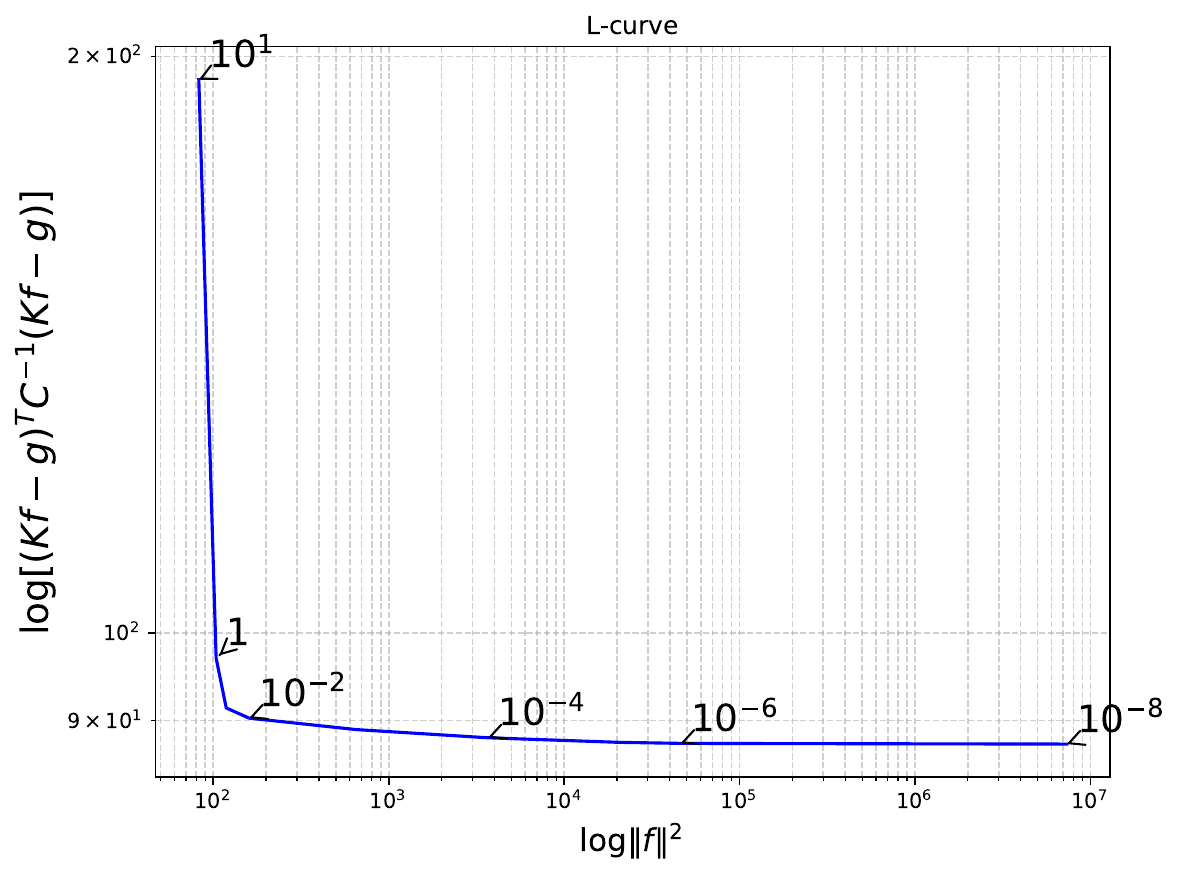}
    \caption{L-curve for determining the regularization parameter $\alpha$ in Tikhonov method for Toy model {\uppercase\expandafter{\romannumeral2}}}
    \label{fig_L_curve_2_TR}
\end{figure}

In order to examine the applicability and stability of the methods mentioned before, we adopt a more complicated test function serving as the true solution, given by:
  \begin{align}
      f(x)=(\sqrt[3]{x}+x^2)e^{-x^2+x}+\text{sin}(e^{\frac{\pi x}{2}}), \quad x\in[-2,1.5].\label{eq:toymodel2}
  \end{align}
The numerical model is generated by the following parameters:
  \begin{align}
      &x_{\text{min}}=-2,\quad x_{\text{max}}=1.5,\quad \Delta x=0.05, \notag\\
      &\lambda_{\text{min}}=-15,\quad \lambda_{\text{max}}=15, \quad \Delta\lambda=0.5.
  \end{align}
The construction of errors for $g(\lambda)$ 
follow the approach used in Toy model \uppercase\expandafter{\romannumeral1}. Similarly, the true solution $f(x)$ in momentum space and its corresponding error-contaminated data $g^{\delta}(\lambda)$ are shown in Fig.~\ref{fig_fake_data2}. The parameters of each method are set as follows: 
\begin{itemize}
    \item Tikhonov: $\alpha_{\text{TR}}=0.1$(determined from the L-curve in Fig.~\ref{fig_L_curve_2_TR}).
    \item Backus-Gilbert: $\alpha_{\text{BG}}=0.1$.
    \item Bayesian: \\
    (1) Flat distribution for $p(f_0|I)$. (lack of prior knowledge about the initial value $f(x_0)$)\\
    (2) Hyper-prior scale parameter $\eta$=0.05 (see Appendix.~\ref{sec_hyperprior_determination} for details on the parameter selection).
    \item ANN: 4 hidden layers of 480 neurons with the Swish activation function $\text{Swish}(t)=\frac{t}{1+e^{-t}}$, performing a hybrid approach combining genetic algorithm with gradient descent for parameter optimization.
\end{itemize}

\begin{figure}
    \centering
    \includegraphics[width=1.0\linewidth]{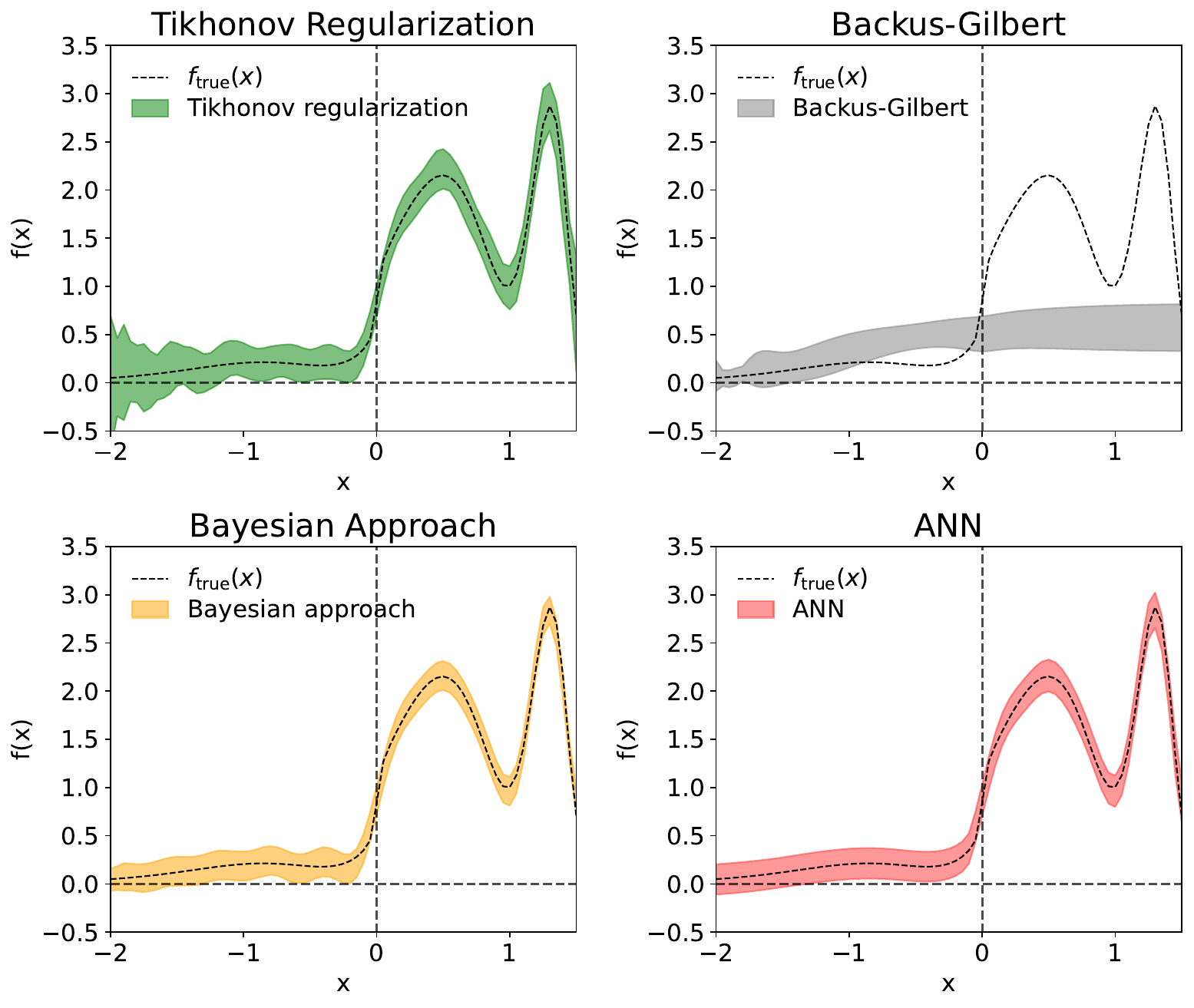}
    \caption{Reconstruction results of Toy model {\uppercase\expandafter{\romannumeral2}} obtained by four methods.}
    \label{fig_Toymodel2_result}
\end{figure}

As a result, Fig.~\ref{fig_Toymodel2_result} presents the reconstruction results from the noisy dataset based on Toy model \uppercase\expandafter{\romannumeral2}.
As can be observed from these figures, the Backus-Gilbert method has completely failed in this complicated case, unable to yield a reconstruction consistent with the behavior of the true solution. 
In contrast, the other three methods still successfully reproduce results that are in good agreement with the true solution. However, both the Bayesian and Tikhonov methods exhibit spurious oscillations in the region $x \in [-2, -0.5]$. This behavior arises because these approaches employ regularization to suppress high-frequency oscillations caused by missing information, and the residual oscillations reflect incomplete suppression. Notably, the Bayesian approach yields weaker spurious oscillations than the Tikhonov regularization. This improvement arises because the prior distribution introduced in Eq.~(\ref{eq_Bayesprior}) enforces smoothness not only on the reconstructed solution itself but also on its first order derivative, thereby enhancing the suppression of high-frequency artifacts.

Meanwhile, the ANN achieves the smoothest reconstruction among all the methods. 
This outcome is not surprising, as the ANN employed here contains a large number of neurons, corresponding to higher parameter freedom and an inherently smoother effective regularization. 
Nevertheless, this excellent performance comes at a substantial cost: the dramatically increased  number of neurons compared to that in Toy Model~\uppercase\expandafter{\romannumeral1}.
In addition to the increased network scale, the architecture employs the Swish activation function, which provides enhanced nonlinear representational capacity and thus further contributes to the overall model complexity. 
This observation clarifies why applying ANNs to inverse problems with intricate solution structures, such as those involving multi-peak spectral function reconstruction, can be particularly demanding. 
Nonetheless, given sufficient computational resources, ANNs retain substantial potential for tackling such complex inverse problems.

\subsection{Real data from lattice QCD simulation}
\begin{figure}
    \centering
    \includegraphics[width=1\linewidth]{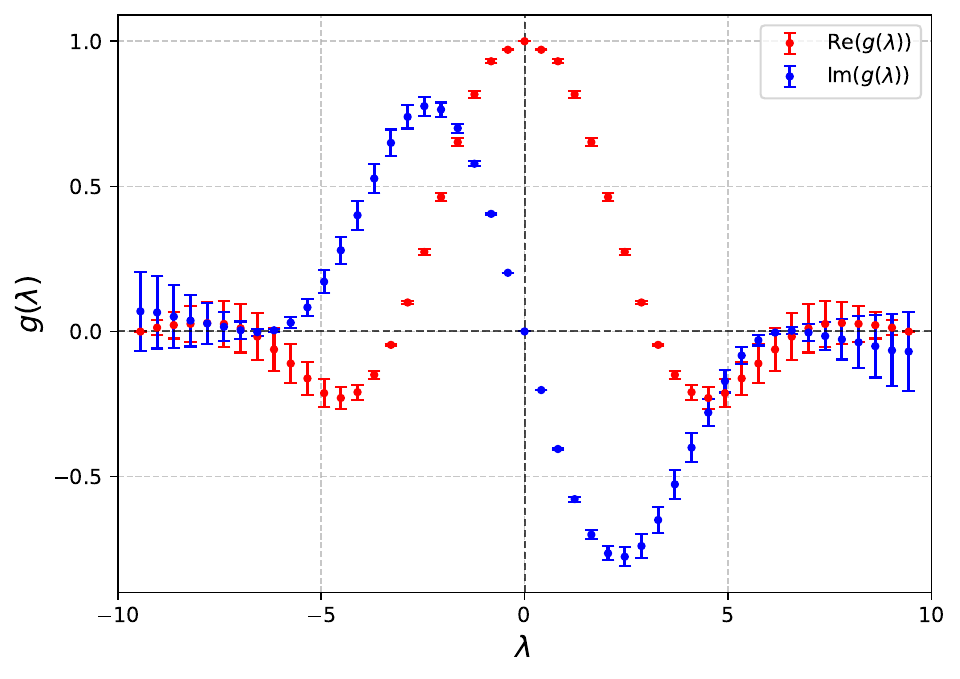}
    \caption{Numerical results for the pion quasi DA from the lattice QCD calculation }
    \label{fig_quasiDA_data}
\end{figure}

\begin{figure}
    \centering
    \includegraphics[width=0.9\linewidth]{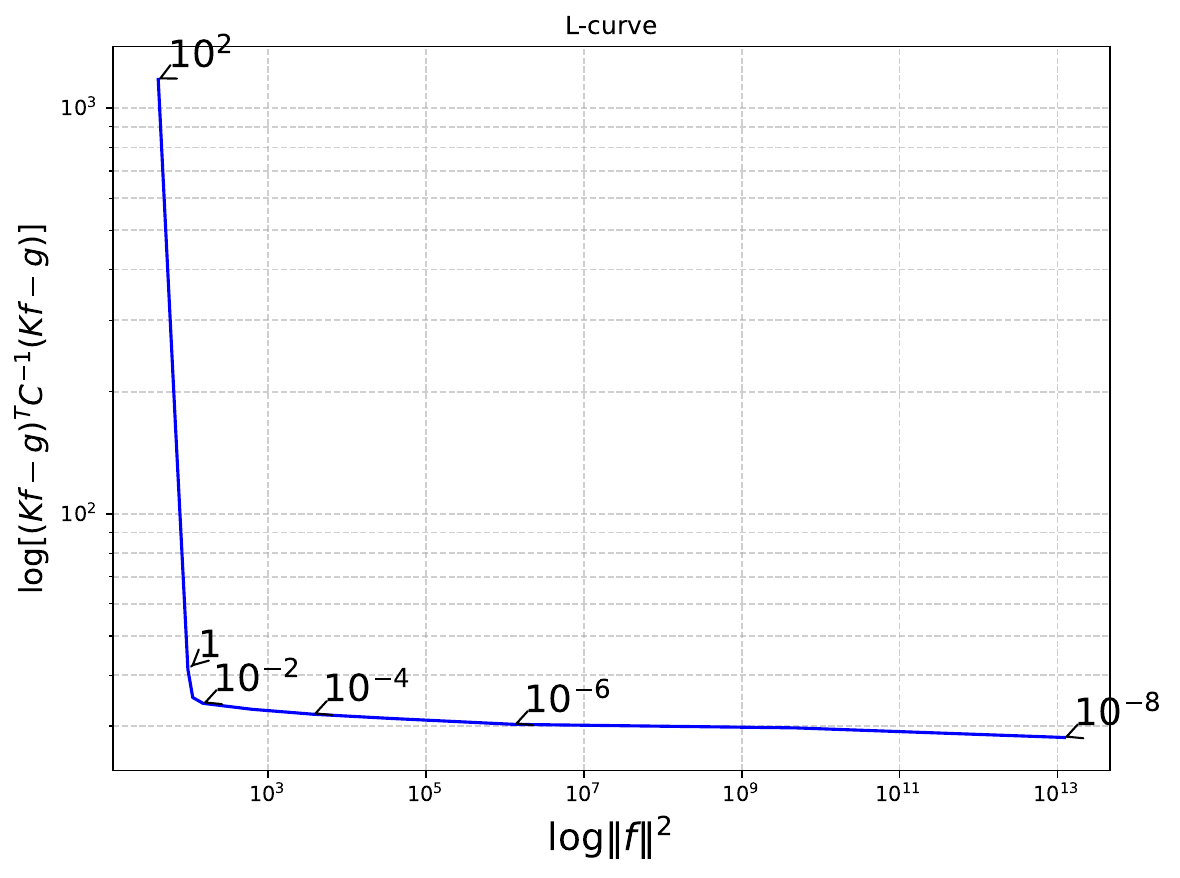}
    \caption{L-curve for determining the regularization parameter $\alpha$ in Tikhonov method with real lattice QCD data input.}
    \label{fig_L_curve_lattice}
\end{figure}

The numerical tests of the four methods on the toy models in the previous subsection have demonstrated their levels of efficacy, respectively. We now apply these four approaches to reconstruct the momentum-space quasi distribution amplitude, using real lattice QCD data of quasi-DA from the Lattice Parton Collaboration (LPC) \cite{LatticeParton:2022zqc}. Fig.~\ref{fig_quasiDA_data} displays the real quasi distribution amplitude in coordinate space computed from lattice QCD. The parameter settings of each method are given by: 
\begin{itemize}
    \item Tikhonov: $\alpha_{\text{TR}}=0.5$(determined from the L-curve in Fig.~\ref{fig_L_curve_lattice}).
    \item Backus-Gilbert: $\alpha_{\text{BG}}=1.2$.
    \item Bayesian: Follow Toy model \uppercase\expandafter{\romannumeral1}. 
    \item ANN: Follow Toy model \uppercase\expandafter{\romannumeral1}.
\end{itemize}

\begin{figure}
    \centering
    \includegraphics[width=1\linewidth]{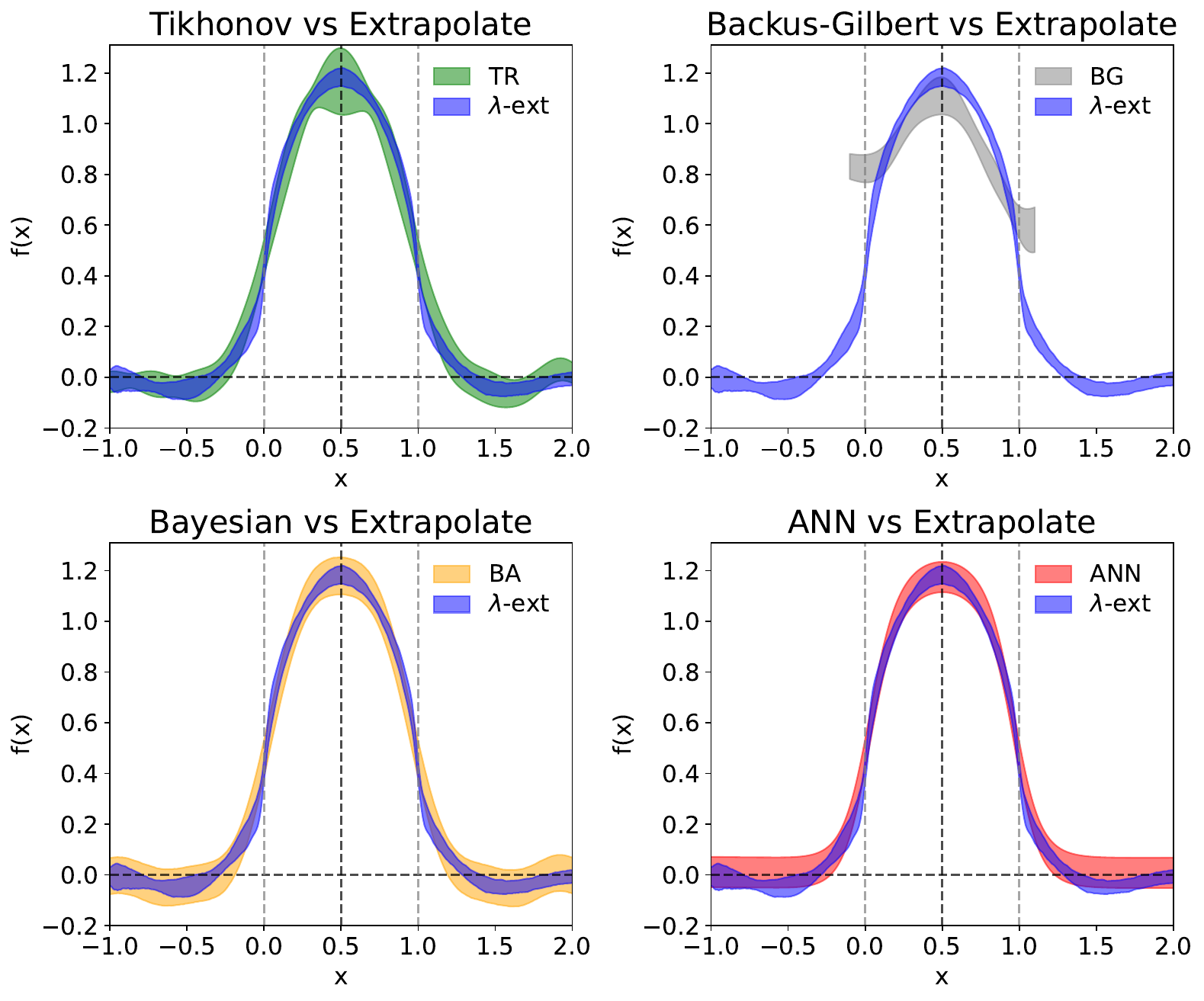}
    \caption{Reconstruction results obtained by the four methods using real lattice data as input, presented in comparison to the result generated by the $\lambda$-extrapolation approach.}
    \label{fig_lattice_result}
\end{figure}

\begin{figure}
    \centering
    \includegraphics[width=1\linewidth]{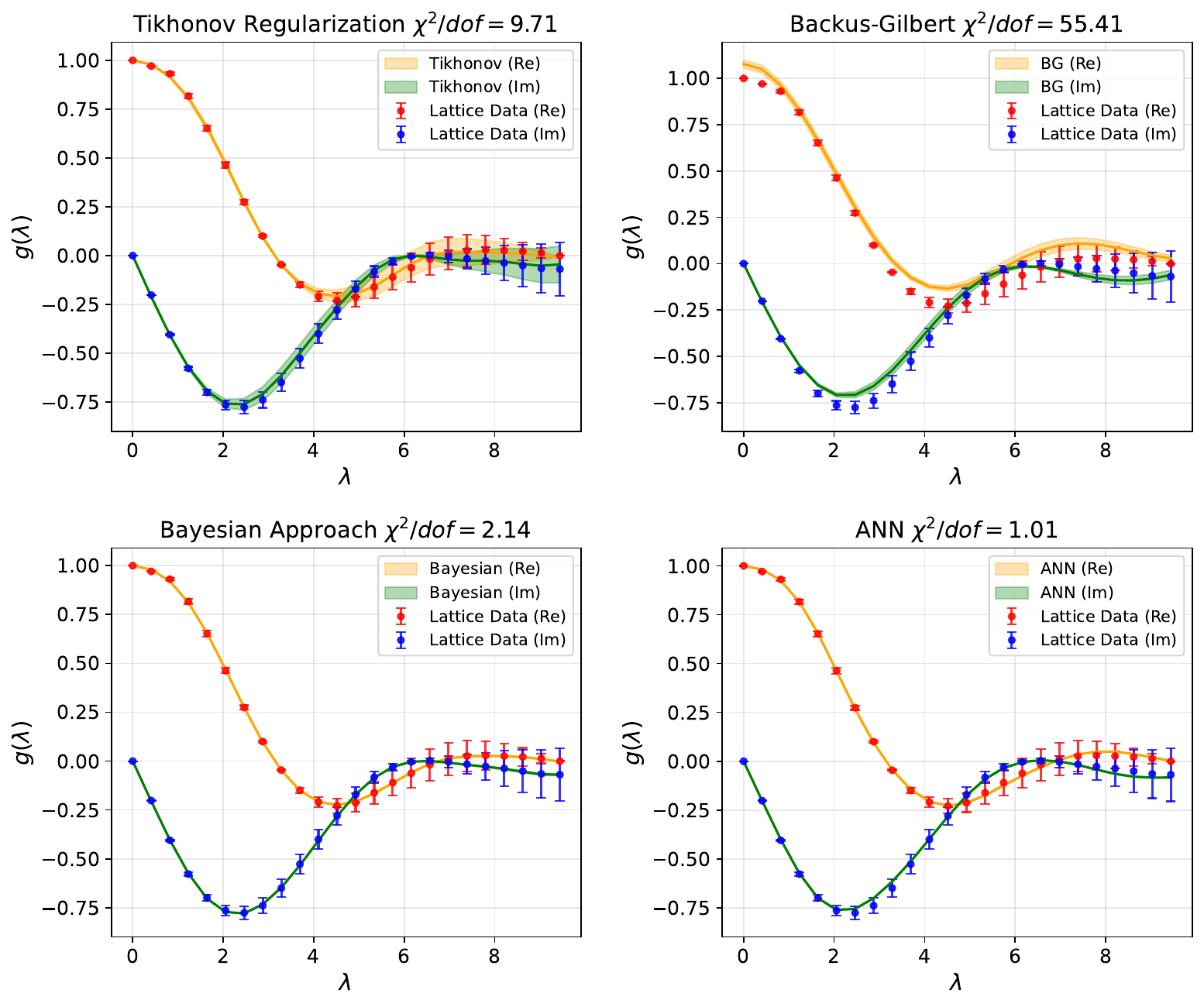}
    \caption{Reconstructed $g(\lambda)$ in coordinate space, obtained by Fourier-transforming the pion quasi-DA from four methods, compared with actual lattice data.}
    \label{fig_lattice_datafit}
\end{figure}

The reconstruction results obtained using the four different methods are presented in Fig.~\ref{fig_lattice_result},
It can be observed that the Tikhonov regularization, Bayesian approach, and ANN again yield largely consistent results in the physical region, which also align well with that derived from the $\lambda$-extrapolation approach. To verify that the reconstructed results in momentum space indeed correspond to the given coordinate-space data, we performed a Fourier transform of the reconstructed solutions back into coordinate space and compared them with the original lattice data.
As shown in Fig.~\ref{fig_lattice_datafit}, the comparison demonstrates that the three methods (excluding the BG method) successfully recover the true solution from the initial lattice data. To better quantify the consistency between the reconstructed data and the original lattice data, we define a chi-squared measure as $\chi^2 = \sum_{i=1}^{n_\lambda} \left( \frac{g_i - \hat{g}_i}{\delta g_i} \right)^2,$ which provides a quantitative evaluation of the deviation between the reconstructed results and the original data.

In contrast, the result from the BG method is noticeably inferior to those of the other methods, as it only captures the behavior in the interval $x\in[0.25, 0.75]$. Combining with the BG's performance on both two toy models and real lattice QCD data, it appears that the BG method is suitable for recovering local features of functions but is inadequate for reconstructing global behavior. This also explains why the Bayesian reconstruction (BR) method~\cite{PhysRevLett.111.182003} is more commonly employed in spectral analyses that primarily focus on identifying the positions of peaks. 

Therefore, based on all the test results from both Toy Model~I, Toy Model~II, and lattice quasi distribution, a consistent picture emerges. For inverse problems where the overall shape of the underlying distribution is of central interest, such as DAs or PDFs, the Bayesian approach with GRW prior and the feedforward ANN demonstrate clear advantages, offering more stable and physically reliable reconstructions. In particular, the ANN exhibits excellent performance for cases characterized by smooth, single-peaked structures, achieving high accuracy with relatively modest computational cost. In this section, we no longer compare the behavior of the $\lambda$-extrapolation approach in coordinate space. Since this method effectively completes the missing large-$\lambda$ data and thereby avoids the ill-posed inverse problem, its reconstructed coordinate-space data are, by construction, consistent with the original lattice data within the statistical uncertainties. 
It should be noted that for all inverse methods, the reconstructed results in coordinate space generally exhibit smaller uncertainties than the original data. From the perspective of the ANN approach, this behavior is quite natural: an inverse method essentially searches for an optimal reconstructed function that best fits a set of noisy input data. During this process, the correlations among different coordinate-space data points serve as crucial information that stabilizes the solution. Consequently, when the reconstructed solution is transformed back to coordinate space, its associated uncertainty naturally becomes smaller than that of the original lattice data. Nevertheless, this reduction in uncertainty does not imply an actual increase in the physical information content; rather, it reflects the smoothing and regularization constraints imposed by the inverse reconstruction.

To examine the robustness limits of the inverse-problem approach under varying data quality, we systematically investigate the breakdown point of the reconstruction as the $\lambda$-range is truncated. The detailed results and analysis are presented in Appendix.~\ref{sec_Truncation_test}.

\section{Summary and outlook}\label{sec_summary}
In this work, we briefly review the definition of ill-posed inverse problems and point out that the limited discrete Fourier inversion problem constitutes an ill-posed inverse problem, which satisfies the existence and uniqueness of a solution but fails to meet the requirement of stability. To address this ill-posed inverse problem, we systematically investigate four reconstruction methods: Tikhonov regularization, the Backus-Gilbert method, the Bayesian approach with GRW, and the feedforward artificial neural networks. These methods are tested on both toy models and actual lattice QCD data. Our numerical results demonstrate that all methods except for the Backus-Gilbert approach can successfully reconstruct the quasi distribution in momentum space. Among them, the Bayesian and ANN methods exhibit superior stability and accuracy.  Meanwhile, the results obtained from these methods also show a good consistency with that derived via the $\lambda$-extrapolation approach.

Our study highlights two complementary strategies for addressing the limited discrete Fourier inversion problem. We demonstrate that both the $\lambda$-extrapolation approach and the inverse problem based approach can achieve reliable reconstructions when the input data are sufficiently precise. Each method, however, has its own strengths and limitations. The $\lambda$-extrapolation approach provides trustworthy uncertainty estimates when data at large $\lambda$ ($\lambda_{\text{max}} > 10$) are precise enough, but extrapolations in regions where $\lambda$ is not sufficiently may introduce additional systematic uncertainties. In contrast, the inverse methods can yield reliable reconstruction results when the data are accurate in the intermediate $\lambda$ ($\lambda_{\text{max}} \sim 6-10$) region. Given that lattice data often exhibit rapidly decaying signals at large $\lambda$, inverse methods particularly the Bayesian and ANN approaches, may offer a more cost-effective way to reconstruct momentum-space distributions.

Therefore, our work suggest that researchers, whether working within the LaMET framework or the pseudo-distribution method can treat the limited inverse discrete Fourier transform (L-IDFT) as a well-defined inverse problem that can be properly addressed. Unlike the inverse problems encountered in lattice spectral analyses or in reconstructing distributions from OPE moments, this is a moderately tractable ill-posed problem, which can be effectively managed when the data quality at large nonlocal separations is sufficiently good.
However, when the data suffer from excessively short $\lambda$ truncation or insufficient precision, the inverse FT becomes ill-conditioned to the extent that meaningful reconstruction is no longer feasible. In such cases, improving the precision of the lattice data itself is a prerequisite rather than refining the inversion method. In contrast, given data with sufficiently large $\lambda$ coverage and adequate precision, one can select an appropriate reconstruction strategy based on the characteristics of the data. A more comprehensive approach would involve employing multiple methods simultaneously to provide a robust estimate of the systematic uncertainties.

\section{Acknowledgments}

We thank Andreas Sch\"afer, Yong Zhao and Xiangdong Ji for valuable discussions and constructive suggestions on this work.
We also thank Ting Wei and Xiong-Bin Yan for valuable  discussions in mathematics.   We thank the LPC collaboration for providing us the lattice data of pion distribution amplitudes \cite{LatticeParton:2022zqc}. This work is supported in part by Natural Science Foundation of China under grant No. 12205106, 12375069 and 12335003. M.~H.~C. is supported by the National Science Centre (Poland) grant OPUS No.\ 2021/43/B/ST2/00497. Q.A.Z is supported by the Fundamental Research Funds for the Central Universities. A.S.X is supported the Fundamental Research Funds for the Central Universities under No.~lzujbky-2023-stlt01.



\appendix 
\begin{widetext}
\section{A proper $\lambda$-truncation}\label{sec_Truncation_test}
While the effectiveness of the inverse-problem-based approach for solving the limited inverse discrete Fourier transform (L-IDFT) has been established in the main text, this appendix investigates its limitations by systematically examining how the reliability of the reconstruction degrades as the data quality diminishes. Specifically, we determine the threshold in the $\lambda$-truncation such that the inverse-problem approach fails to yield a trustworthy result. This analysis provides practical guidance on the necessary precision and range of lattice data required to guarantee a reliable reconstruction of PDFs or DAs using these methods. Based on this framework, we employ Toy Model \uppercase\expandafter{\romannumeral1} as the test case and the Tikhonov method and ANN as the reconstruction platforms. The numerical tests are designed:
\begin{itemize}
    \item The data spacing $\Delta\lambda=0.6$ is fixed. Starting from a maximum $\lambda_{\text{max}}=9.6$, we progressively truncate the data range to smaller $\lambda_{\text{max}}$ values.
\end{itemize}

Fig.~\ref{fig_data_cutoff} shows the data in coordinate space under a decreasing $\lambda_{\text{max}}$, while the corresponding momentum-space reconstructions from the Tikhonov and ANN methods are provided in Fig.~\ref{fig_Tik_cutoff} and Fig.~\ref{fig_ANN_cutoff}, respectively. It can be observed that as $\lambda$ gradually decreases, the quality of the reconstructions from both the Tikhonov and ANN methods degrades markedly. This implies that when $\lambda_{\text{max}}$ becomes too small, the amplification of the condition number causes the instability inherent to ill-posed problems to grow so severe that the problem can no longer be reliably solved. 
In the case of the ANN method, the reconstruction error starts to increase rapidly and becomes uncontrollable once $\lambda_{\text{max}} < 6$. 
In contrast, for explicit-regularization-based methods such as the Tikhonov approach, the reconstructed results remain smooth and the uncertainties appear well-controlled even for $\lambda_{\text{max}} < 6$; however, a clear deviation from the true solution emerges, indicating the onset of an uncontrollable systematic bias. 
The distinct behaviors of the two methods in the low$\lambda_{\text{max}}$ regime originate from their fundamentally different solving strategies:
the ANN, which determines the optimal solution through loss-function minimization, is intrinsically more sensitive to numerical instabilities, 
whereas the Tikhonov regularization tends to over rely on the regulator when the input data become insufficient, thus introducing systematic deviations. 

Overall, regardless of the specific form of breakdown, these observations indicate that the applicability of inverse methods for solving the L-IDFT problem is inherently limited. 
Based on our numerical tests—and considering that realistic physical solutions are typically more complex than those in the toy models—we recommend that, when extracting parton distribution functions (PDFs) or distribution amplitudes (DAs) within the LaMET or pseudo-distribution frameworks, the input data should extend to at least $\lambda_{\text{max}} > 6$ to ensure a reliable and effective reconstruction. 
Below this threshold, even highly precise data in coordinate space are insufficient to yield a trustworthy momentum-space solution.



\begin{figure}[!htbp]
    \centering
    \begin{minipage}{1.0\linewidth}
        \centering
        \includegraphics[width=\linewidth]{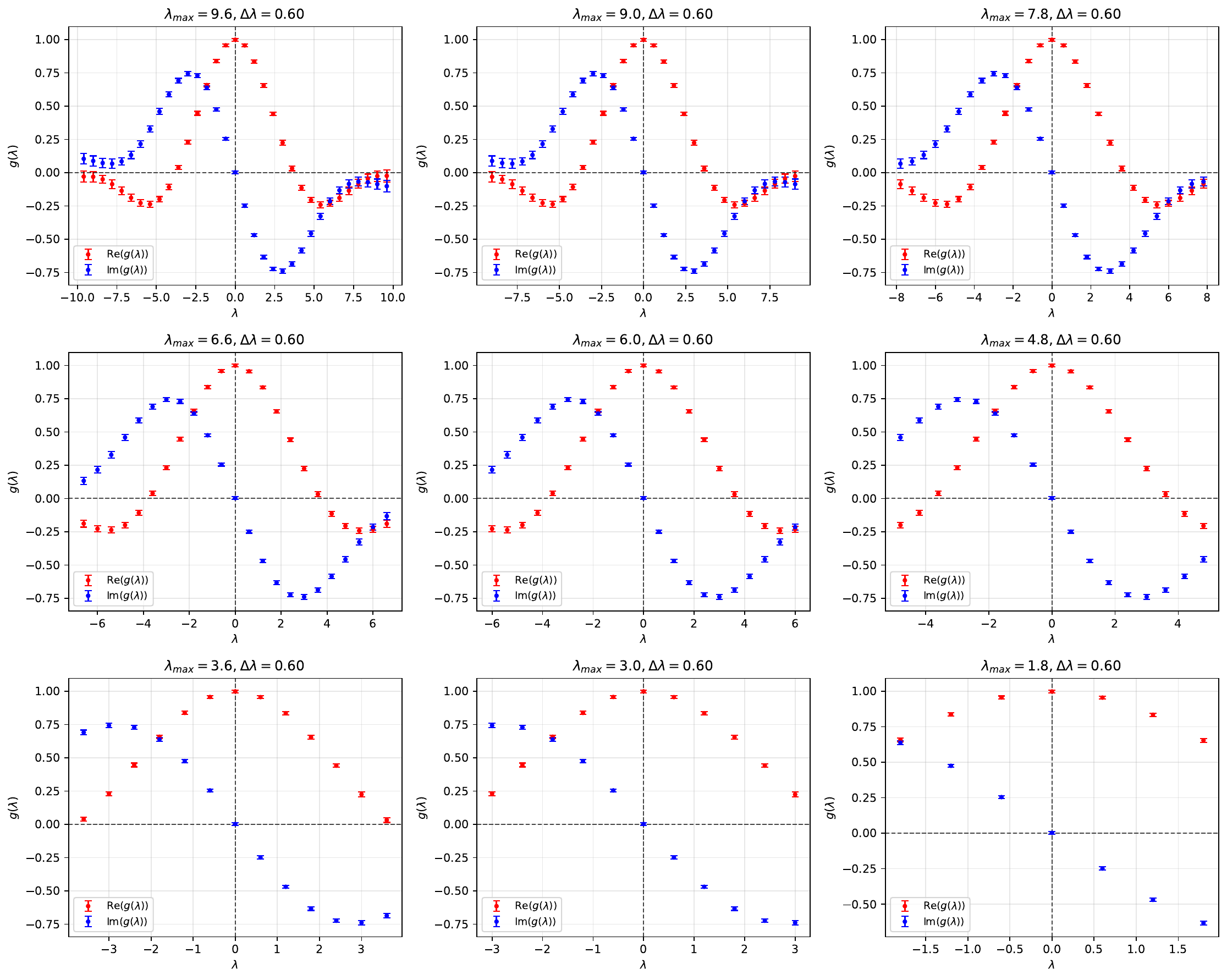}
        \caption{Coordinate space data $g(\lambda)$ with different $\lambda_{\text{max}}$.}
        \label{fig_data_cutoff}
    \end{minipage}
\end{figure}

\begin{figure}[!htbp]
    \centering
    \begin{minipage}{1.0\linewidth}
        \centering
        \includegraphics[width=\linewidth]{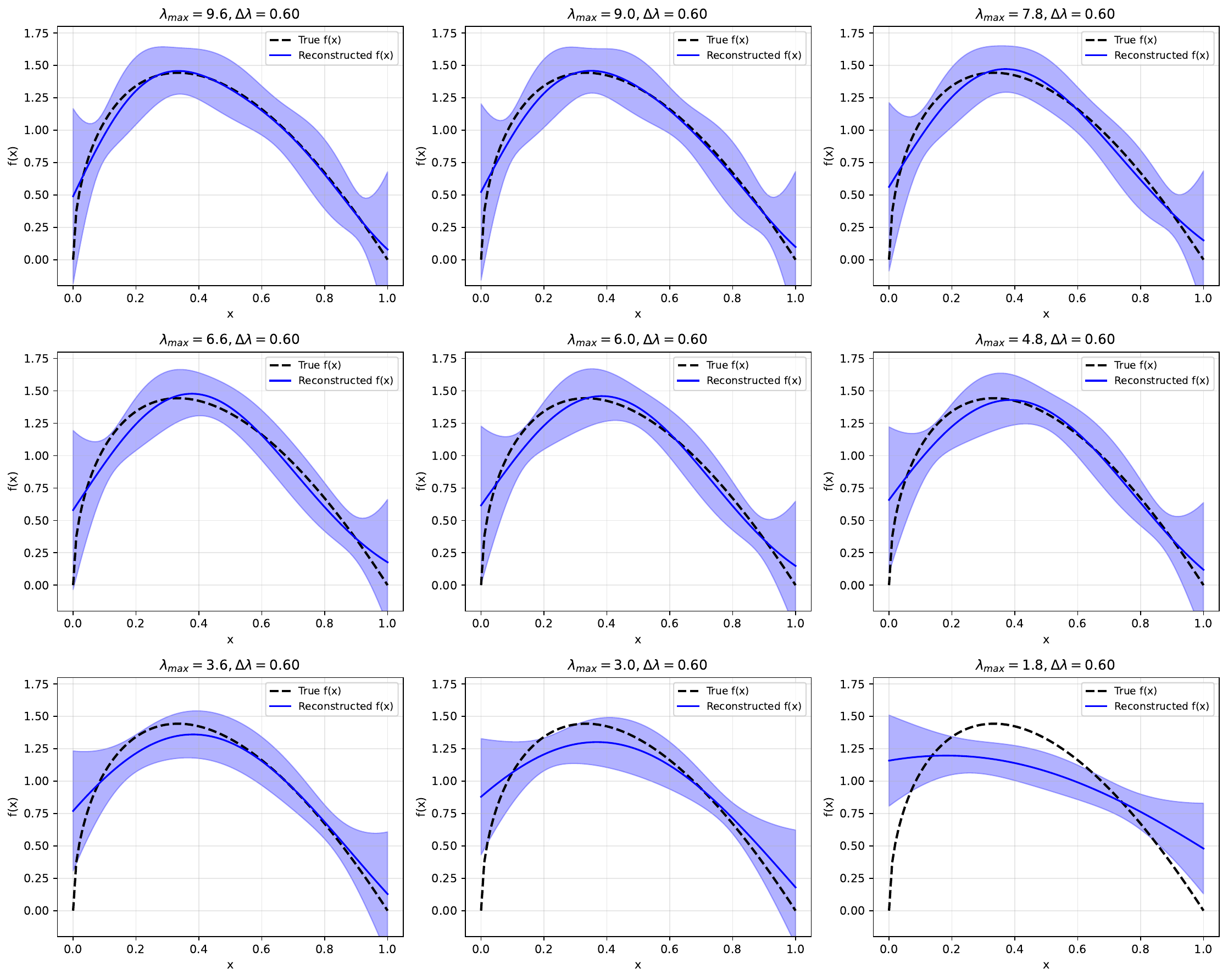}
        \caption{Momentum space reconstructions using the Tikhonov method under different $\lambda_{\text{max}}$ values.}
        \label{fig_Tik_cutoff}
    \end{minipage}
\end{figure}
\clearpage
\begin{figure}[!htbp]
    \centering
    \begin{minipage}{1.0\linewidth}
        \centering
        \includegraphics[width=\linewidth]{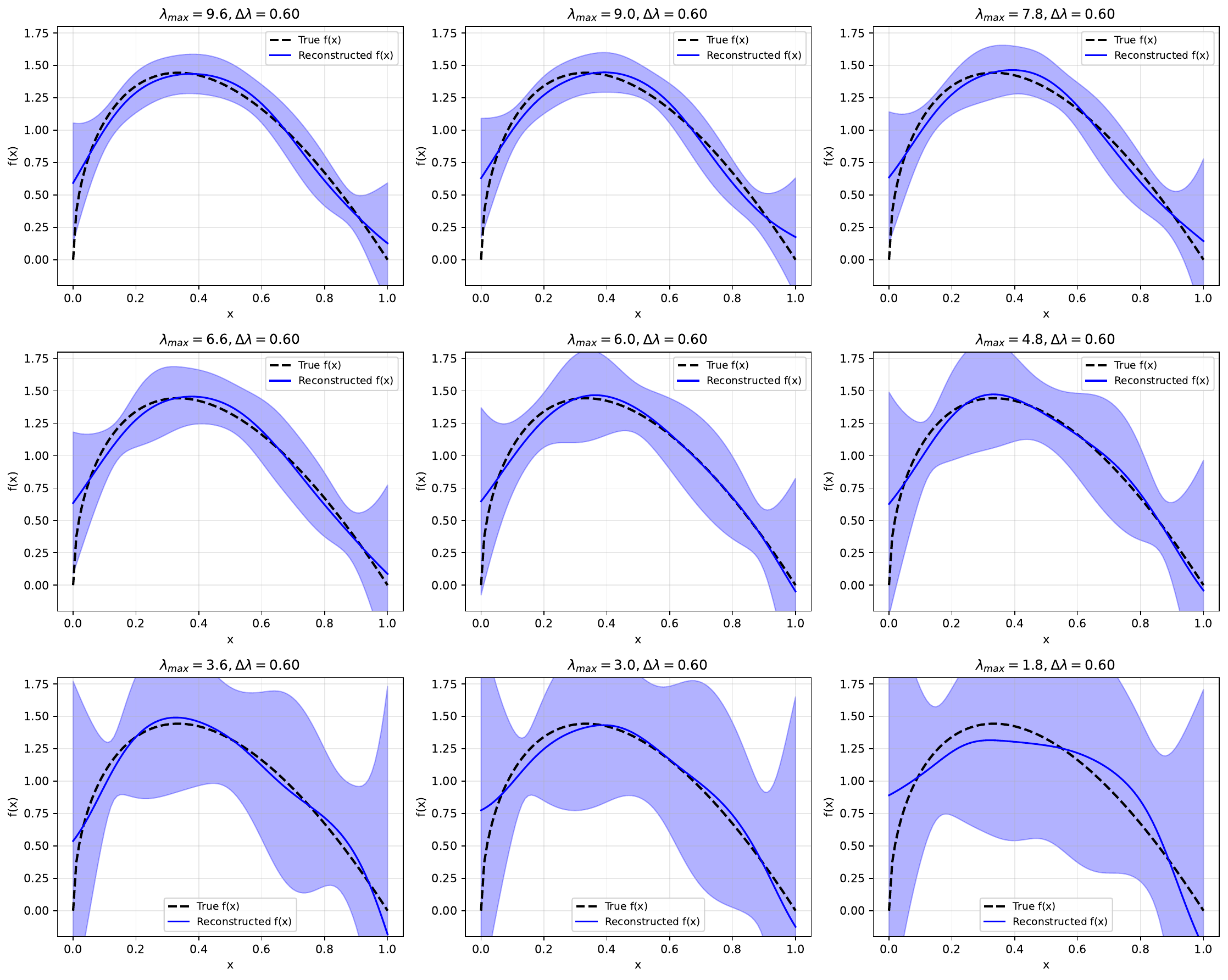}
        \caption{Momentum space reconstructions using the ANN method under different $\lambda_{\text{max}}$ values.}
        \label{fig_ANN_cutoff}
    \end{minipage}
\end{figure}

\section{Determination of scale parameter $\eta$ in Bayesian approach}\label{sec_hyperprior_determination}

The inversion results for both Toy Model I and Toy Model II (Fig. \ref{fig:toy1-eta} and \ref{fig:toy2-eta}, respectively) demonstrate robustness against the specific choice of $\eta$. Both models exhibit a broad stability plateau: for Toy Model I, solutions are stable around $\eta = 0.01$, and for Toy Model II, around $\eta = 0.05$. Outside this plateau, an excessively large $\eta$ leads to significant errors due to insufficient regularization, while an overly small $\eta$ causes deviation from the true solution. We estimate the associated systematic error by perturbing $\eta$ within these stable regions. Based on the physical analogy between Toy Model I and the asymptotic behavior of the $\pi$ DA, we fix $\eta=0.01$ for Toy model I and real lattice QCD data.

\begin{figure}[!htbp]
    \centering
    \begin{minipage}{1.0\linewidth}
        \centering
        \includegraphics[width=\linewidth]{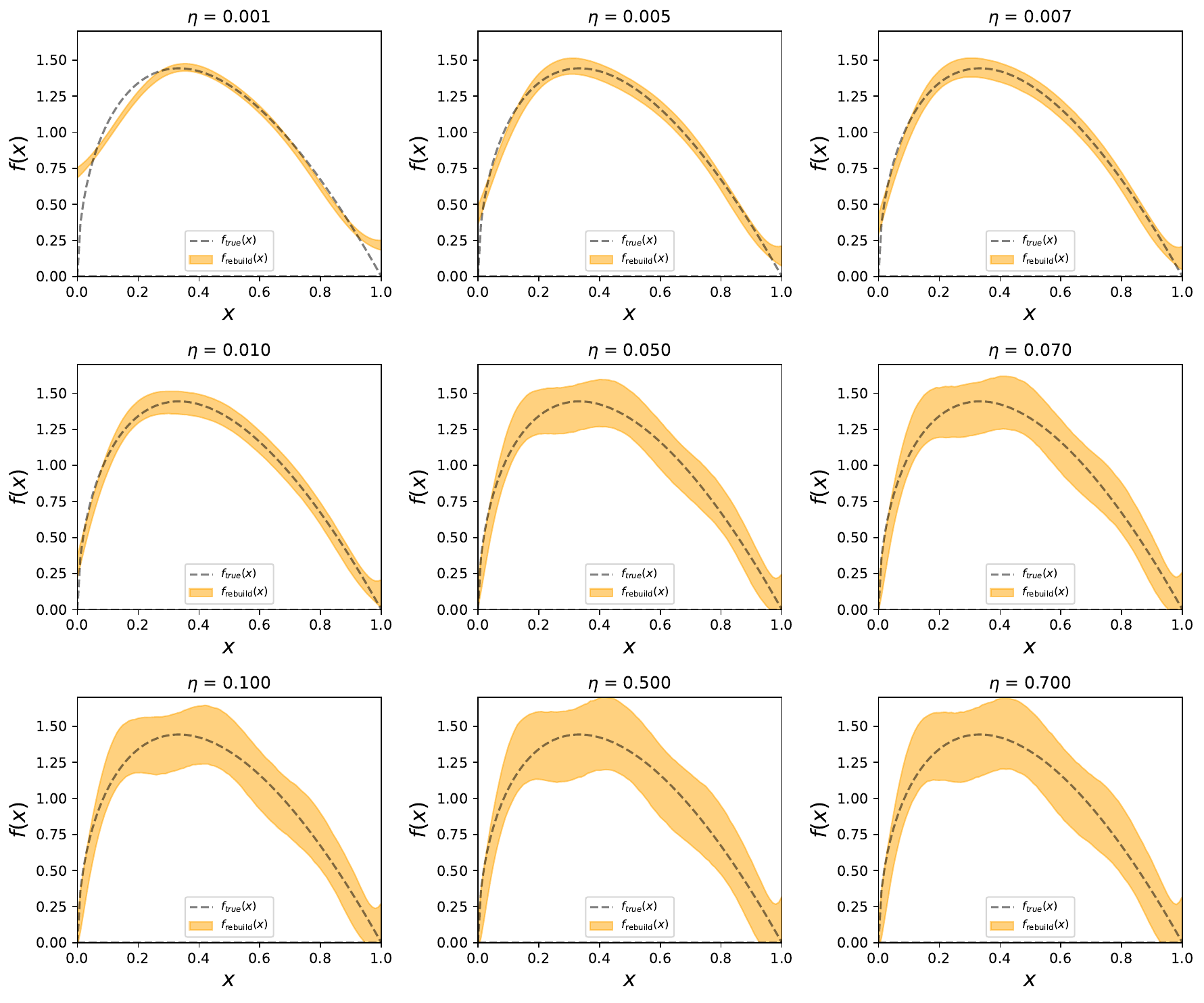}
        \caption{Reconstruction of Toy Model \uppercase\expandafter{\romannumeral1} using the Bayesian method with varying $\eta$ values.}
        \label{fig:toy1-eta}
    \end{minipage}
\end{figure}
\clearpage
\begin{figure}[!htbp]
    \centering
    \begin{minipage}{1.0\linewidth}
        \centering
        \includegraphics[width=\linewidth]{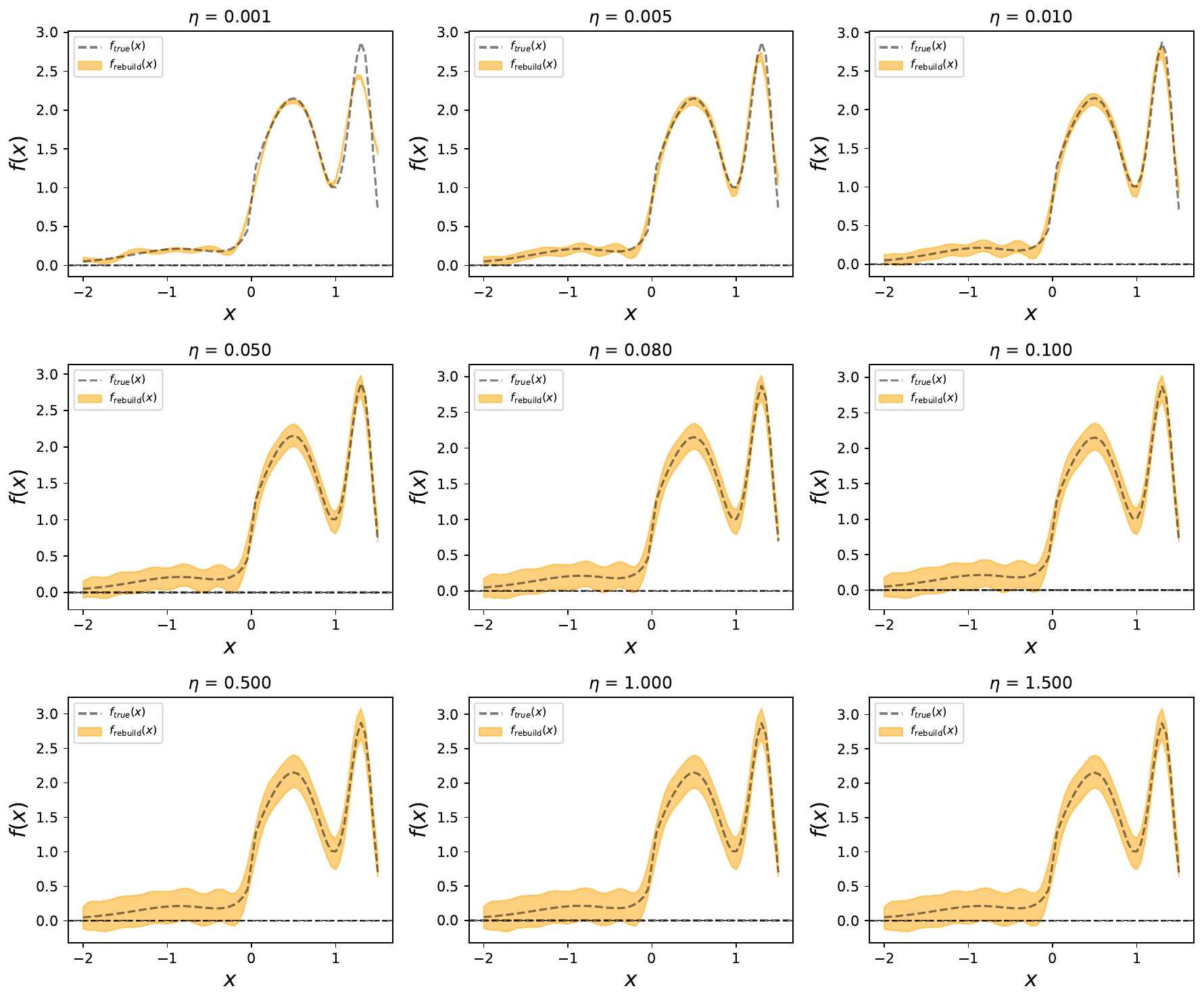}
        \caption{Reconstruction of Toy Model \uppercase\expandafter{\romannumeral2} using the Bayesian method with varying $\eta$ values.}
        \label{fig:toy2-eta}
    \end{minipage}
\end{figure}

\end{widetext}

\clearpage
\bibliography{Refs}
\bibliographystyle{apsrev4-1}

\end{document}